\documentclass[lineno]{jfm}
\usepackage{graphicx}
\usepackage{lineno}
\usepackage{epstopdf, epsfig}
\usepackage{amsmath}
\usepackage{cases}
\usepackage[mathscr]{euscript}
\usepackage{IEEEtrantools}
\usepackage[section]{placeins}
\usepackage{xcolor}

\shorttitle{Receptivity of jets due to acoustic excitations}
\shortauthor{B. Li, S. Zhang and B. Lyu}
\title{Receptivity of a supersonic jet due to acoustic excitations near the nozzle lip}

\author{Binhong Li\aff{1}, Sicheng Zhang\aff{1}
   and Benshuai Lyu\aff{1,2}
\corresp{\email{b.lyu@pku.edu.cn}}
}

\affiliation{\aff{1}State Key Laboratory of Turbulence and Complex
    Systems, College of Engineering, Peking University,  5 Yiheyuan Road,
    Haidian District, Beijing 100871, China
    \aff{2}Laoshan Laboratory, Qingdao 266100, PR China
}
\begin{document}
\maketitle
\begin{abstract}
In this paper, we develop an analytical model to investigate the generation of instability waves triggered by the upstream acoustic forcing near the nozzle lip of a supersonic jet. This represents an important stage, i.e. the jet receptivity, of the screech feedback loop. The upstream acoustic forcing, resulting from the shock-instability interaction, reaches the nozzle lip and excites new shear-layer instability waves. To obtain the newly-excited instability wave, we first determine the scattered sound field due to the upstream forcing using the Wiener-Hopf technique, with the kernel function factored using asymptotic expansions and overlapping approximations. Subsequently, the unsteady Kutta condition is imposed at the nozzle lip, enabling the derivation of the dispersion relation for the newly-excited instability wave. A linear transfer function between the upstream forcing and the newly-excited instability wave is obtained. We calculate the amplitude and phase delay in this receptivity process and examine their variations against the frequency. The phase delay enables us to re-evaluate the phase condition for jet screech and propose a new frequency prediction model. The new model shows improved agreement between the predicted screech frequencies and the experimental data compared to classical models. It is hoped that this model may help in developing a full screech model. 
%We show that the predicted sound directivity is in good agreement with experimental data. In particular, this model shows the strongest noise emission close to the upstream direction but starts to rapidly decay at $180^\circ$, which is in accordance with experimental results; this suggests that the noise from a single shock cell is far from of the monopole type as assumed in the classical Powell's model. 
%We find that the noise directivity is very sensitive to the local growth rate of the instability waves, which may enable an explanation for the discrepancies observed between different
\end{abstract}

%\section{Nomenclature}

%{\renewcommand\arraystretch{1.0}
%\noindent\begin{longtable*}{@{}l @{\quad=\quad} l@{}}
%$A$  & amplitude of oscillation \\
%$a$ &    cylinder diameter \\
%$C_p$& pressure coefficient \\
%$Cx$ & force coefficient in the \textit{x} direction \\
%$Cy$ & force coefficient in the \textit{y} direction \\
%c   & chord \\
%d$t$ & time step \\
%$Fx$ & $X$ component of the resultant pressure %force acting on the vehicle \\
%$Fy$ & $Y$ component of the resultant pressure %force acting on the vehicle \\
%$f, g$   & generic functions \\
%$h$  & height \\
%$i$  & time index during navigation \\
%$j$  & waypoint index \\
%$K$  & trailing-edge (TE) nondimensional angular %deflection rate
%\end{longtable*}}

\section{Introduction}
\label{Sec:introduction}
%\ccc{Introduction to jet screech and Powell's pioneering work.} 
Screech is a phenomenon characterized by discrete acoustic tones in imperfectly expanded supersonic jet flows \citep{Raman_1999, 2019EDINGTON}. Its high intensity can lead to sonic fatigue and severe damage to aircraft structures, emphasizing the importance of understanding its underlying mechanism. In the pioneering work by \citet{19533Powell}, he proposed the well-known feedback-loop mechanism consisting of four stages: the downstream propagation of instability waves, the interaction between shock and instability waves, the upstream propagation of acoustic waves, and the jet receptivity at the nozzle lip. 
Powell further introduced the gain and phase conditions necessary for establishing the screech.  These two conditions laid the foundation of following jet screech research.

%\ccc{Frequency prediction model for jet screech.} 
The screech frequency, herein defined as $f$, has gained much attention in early research. Powell suggested that it can be calculated by assuming constructive interference at $180^\circ$ between shock-cell sources, i.e.
\begin{equation}
f=\frac{U_{c}}{s(1+M_{c})},
\label{equ_screech f}
\end{equation}
where $s$, $U_{c}$, and $M_{c}$ are the shock cell spacing, the convection velocity, and the convective Mach number of the instability waves, respectively. Note, strictly speaking, (\ref{equ_screech f}) is not a traditional phase condition in the sense that the overall phase change during the four stages of screech is assumed to be $2N\pi$, where $N$ is an integer. Nonetheless, the predicted results using (\ref{equ_screech f}) demonstrated satisfactory agreement with experimental data regarding rectangular nozzles.  
Subsequently,~\citet{1986Tam_Proposed} proposed the weakest link theory, suggesting screech as the limit of broadband shock-associated noise (BBSAN) when the observer angle approaches $180^\circ$.  Based on these two models, new frequency prediction formulas were developed to include effects of nozzle geometry, including rectangular nozzles with different aspect ratios~\citep{Tam_shock_space,Tam_1996_JoA}, bevelled rectangular nozzles~\citep{tam_bevelled}, and elliptic nozzle~\citep{Tam_1988jsv}. These model predictions exhibited favourable agreement with experimental data.

%\ccc{Mode staging phenomena.} 
However, in round jets, Powell observed sudden changes in the screech frequency with increasing inlet pressure, a phenomenon termed mode staging.  Powell categorized these changes into four distinct modes: A, B, C, and D, with mode A further divided into modes $\rm{A}{1}$ and $\rm{A}{2}$~\citep{M.Merle}. 
It is evident that the aforementioned formulas, which showed continuous frequency predictions against the nozzle inlet pressure, proved inadequate in predicting mode staging. The physical mechanism of mode staging necessitated further investigation. %After Powell, many efforts were made to achieve a more accurate screech frequency prediction. Subsequently,~\citet{1986_Proposed} proposed the weakest link theory suggesting the screech as the limit of the broadband shock-associated noise~(BBSAN) when the observer angle approaches $180^\circ$. In 1999, \citet{Panda_standingwave} discussed the link between screech and hydrodynamic-acoustic~(HA) standing waves, and a new formula was developed. However, these models still seemed unable to predict the abrupt frequency jump. 

%\ccc{Detailed introduction to Li's work on the mode staging and other theories regarding the mode staging.} 
To investigate the mode staging phenomenon, \citet{X.D.Li} introduced the phase condition as an alternative to the constructive interference condition to predict the screech frequency. They conducted comprehensive numerical simulations to thoroughly investigate various parameters regarding the phase condition, including the convection velocity of instability waves, the effective sound source region, and the shock spacing. Their simulation results demonstrated agreement with experimental data when the cycle numbers contained in the feedback loop were changed, suggesting that the phase condition might provide insights into the underlying mechanism of the mode staging phenomenon. Recent years have witnessed a re-evaluation of the phase condition, see for example \citet{Jordan_2018} and \citet{NewC_1}. These new theories emphasized the crucial role played by the guided-jet mode in completing the feedback loop of jet screech, as evidenced by recent publications~\citep{Gojon_modeStaging,Edgington2018,Xiang-Ru_2020}. According to these studies, both the $\rm{A}{1}$ and $\rm{A}{2}$ modes were closed by guided-jet modes rather than free acoustic waves.  Moreover, the guided-jet mode was found to be active in all screech modes~\citep{unifying_theory}, and its excitation can be attributed to interactions between the instability waves and the optimal order as well as suboptimal wavenumber components of the shock structures~\citep{A1A2, unifying_theory, xiangru_Li_2023}. The guided-jet mode due to the interaction between instability waves and the suboptimals is shown to be responsible for closing the feedback loop in multiple stages of jet screech. 
The role of guided-jet modes was also examined under the circumstances of screeching twin jets~\citep{Stavropoulos_recep}, which showed similar trends to those observed in single jets. 
%To the best of our knowledge, one of the best fittings to the experimental data was obtained by~
%\citet{X.D.Li} used the original phase condition shown in~(\ref{phase_condition}) and inferred the value of $N$ in their numerical study. They showed that $N$ differed across various stages, and when  this difference was considered the prediction was in very good agreement with the experimental data. However, the mechanism behind this mode transition and its high sensitivity to the initial conditions are yet to be clarified.

%\ccc{By mentioning the importance of the phase delay in predicting mode staging to introduce the jet receptivity.} 
To invoke the phase condition to predict the mode staging phenomenon, an important step is to determine the phase delay between the upstream forcing and the newly-triggered instability waves. This necessitates a close examination of the receptivity process of the jet screech. Previous investigations conducted relevant research using experimental measurements, numerical simulations, and analytical models.

%\ccc{Experimental study and numerical simulations on the jet receptivity.} 
Early research primarily focused on experimental measurements, which revealed that modifications to the nozzle lip, such as nozzle thickening, could change both the mode and amplitude of screech tones~\citep{1983Norm,PM_SJ_JSV,1997JFM_Raman}. Additionally, an acoustic reflector placed upstream or downstream of the nozzle lip was shown to significantly impact screech amplitude~\citep{NRT_DJW,TDNorum_AIAA, Raman_receptivity}. Moreover, it was found that installing conical reflector surfaces around a round nozzle exit could cause significant changes in the mode staging behaviour~\citep{receptovity_shape}. Recently,~\citet{receptovity_roughness} showed that when the surface roughness of the nozzle lip increases, the screech amplitude decreased or even disappeared.
 These observations underscored the key role played by the jet receptivity in determining the screech amplitude and frequencies.
 
 Besides experimental measurements, several numerical simulations were conducted to investigate jet receptivity.~\citet{Tam_receptivity} explored the effects of nozzle-lip thickness on the intensities of the axisymmetric screech tones. They found that at low supersonic Mach numbers, 
 the changes of the screech amplitude were not significant with a thickened nozzle lip, which aligned with experimental observations~\citep{PM_SJ_JSV}. Recent years have seen an increase in research on jet receptivity. \citet{Karami_receptivity1} defined a transfer function at the nozzle lip between the external disturbance and the initial conditions of the vortical instability in the case of impinging jets. Their findings indicated that the scattering efficiency was related to the nondimensionalized frequency, the azimuthal wavenumber, and the pulse location. Subsequently, \citet{Karami_receptivity2} incorporated the effects of nozzle geometry into their analysis. They noted that an infinite-lipped nozzle supported higher amplitude upstream-propagating waves both inside and outside the jet flow compared to the thin nozzles. 
 \citet{bogey_receotivity1} conducted Large Eddy Simulations (LES) to explore the interaction between upstream-propagating guided jet modes and shear-layer instability waves near the jet nozzle. He observed that the frequency of the most amplified Kelvin-Helmholtz instability wave fell within the narrow frequency bands of the guided-jet mode. The effects of the nozzle thickness were also investigated in \citet{Bogey_receptivity2}, where it was found that a thickened nozzle lip led to an increase in the near-field sound pressure level downstream of the jet nozzle.
 
 %\ccc{Analytic model on receptivity.} 
 Despite numerous experimental and numerical investigations, not many analytical models were developed to model the jet receptivity. \citet{tam_receptivity2} investigated the excitation of instability waves in infinitely-long two-dimensional free shear layer by acoustic waves, which was solved by using the Green's function of the problem. He found that to excite instability waves at moderate subsonic flow Mach numbers, an acoustic beam inclined at an angle of $50^\circ$ to $80^\circ$ to the shear flow is the most effective. Another notable study conducted by~\citet{E.J.Kerschen} examined the jet receptivity due to acoustic excitations and introduced a receptivity coefficient to predict the screech amplitude. However, comprehensive details of this model appeared missing in the open literature. Another study by \citet{SKLELE_receptivity} employed a combined theoretical and computational approach and revealed that upstream-propagating acoustic perturbations could excite instability waves in the jet mixing layer. They further observed that the thickness of the nozzle lip significantly influenced the receptivity process, which accorded with numerous experimental observations. 

%\ccc{Research gap.}
 In spite of numerous studies on jet receptivity, the phase delay in the receptivity process remains unclear. Previous studies often assumed a fixed phase delay such as 0 or $\pi/4$~\citep{Jordan_2018,NewC_1,Stavropoulos_recep,xiangru_Li_2023}. To obtain the correct phase delay, instead of relying on assumed values, an analytical model of the jet receptivity is needed. More specifically, a linear transfer function between the upstream forcing and the newly-generated instability wave is desired. Therefore, the primary objective of this paper is to develop such a model to improve our understanding of jet receptivity.

%\ccc{The methods employed to develop this model and the structure of this paper.} 
This paper is structured as follows: section~\ref{sec:types_paper} provides a detailed analytical derivation of the model, including the formulation of the Wiener-Hopf problem, the decomposition of the kernel function, and the determination of the dispersion relation. A linear transfer function between the upstream forcing and the newly-excited instability waves is obtained. In section \ref{sec:results}, we present our results based on the derived transfer function, where a new formula is proposed to predict the screech frequency. Finally, section \ref{sec:summary} concludes this paper.%section~\ref{section:results and discussion}  shows the prediction of directivity patterns of the fundamental tone and its harmonics. The noise generation mechanism is subsequently discussed. %Additionally, a presume is proposed to interpret mode stagings.
%Conclusions are drawn in section~\ref{section:conclusions}.

\section{Analytical formulation}
\label{sec:types_paper}
\subsection{The Wiener-Hopf equation}
\label{subsection:base assumptions}
Inspired by~\citet{DGCrigh}, who investigated jet-edge tones, we focus on a two-dimensional jet flow of a vortex-sheet type in this paper. To render the model more realistic, the incident acoustic wave takes an asymptotic form generated through the interaction between shock and instability waves, as obtained previously by the authors~\citep{Myown, MyOwn_2}.

\begin{figure}
   \centering
   \includegraphics[width = 0.8\textwidth]{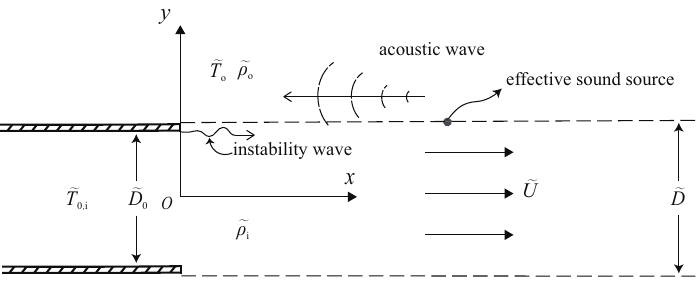}
   \caption{The schematic of the vortex-sheet flow configuration and Cartesian
   coordinates. The origin is fixed at the centre of the nozzle while $x-$ and
  $y-$ axes represent the streamwise and cross-flow coordinates, respectively.}
\label{fig:example1}
\end{figure}

%\ccc{Nondimensionalization.} 
Schematic illustration is shown in figure~\ref{fig:example1}. As we can see, a supersonic jet is issued from the nozzle and takes the form of a vortex sheet. The coordinate axes $(x,y)$ are chosen to be parallel and perpendicular to the nozzle centreline, respectively, while the origin point $O$ is located at the center of the nozzle exit plane. According to \citet{1950Pack}, the base flow is chosen as the fully expanded jet velocity $\tilde{U}$, which is different from the jet exit velocity $\tilde{U}_j$. $\tilde{D}$ is the height of the fully expanded jet. Note that $\tilde{D}$ is generally not equal to the height of the nozzle $\tilde{D}_0$~\citep{tam_diameter}, but is less or greater than $\tilde{D}_0$ depending on whether the jet is underexpanded or overexpanded \citep{tam_machwave}. The densities inside and outside the jet are denoted by $\tilde{\rho}_{\mathrm{i}}$ and $\tilde{\rho}_{\mathrm{o}}$, respectively, while $\tilde{T}_{0,\mathrm{i}}$ and $\tilde{T}_{\mathrm{o}}$ represent the total reservoir temperature and the ambient temperature outside the jet, respectively. 
The nozzle lips, represented by two flat plates in figure~\ref{fig:example1}, are assumed to be semi-infinite with a negligible thickness, extending to $-\infty$ in the $x$ direction. Downstream of the nozzle lip, an acoustic wave, with a speed of sound $\tilde{a}_\infty$ and a frequency of $\tilde{\omega}$, is emitted from a source region and propagates upstream to the nozzle lip, triggering the instability waves.

To simplify the problem, we nondimensionalize velocities and lengths by $\tilde{U}$ and $\tilde{D}$, respectively. In addition, densities and temperatures are nondimensionalized by $\tilde{\rho}_{\mathrm{o}}$ and $\tilde{T}_{\mathrm{o}}$, respectively.  Note that we use the symbols with a tilde to represent dimensional variables, while those without denote non-dimensional variables. For example, after the nondimensionalization, the jet exit velocity is $U_j$ and the base flow takes the form
\begin{equation}
    \boldsymbol{u}_{0}= 
      \begin{cases}
        0, & |y|>1/2, \\
        \mathbf{e}_{x}, &  |y|\leq 1/2,
    \end{cases} 
  \end{equation}
  %\textbf{You need to define $e_x$ and it should be in bold font, Redefineequation 2.8 using $|y| > \frac{D}{2}$to have a compact expression, and use the case environment in latex to do so} 
  where $\mathbf{e}_{x}$ is the unit vector in the $x$ direction.

Considering that the mean pressure outside and inside of the vortex sheet remains the same when a perfect gas is assumed, the density ratio between the outside and inside of the vortex sheet, here defined as $\sigma$, is equal to $M_{\mathrm{o}}^2 / M_{\mathrm{i}}^2$. Here, $M_{\mathrm{o}}$ and $M_\mathrm{i}$ are defined as $1/c_{\mathrm{o}}$ and $1/c_\mathrm{i}$, respectively, where $c_{\mathrm{o}}$ and $c_\mathrm{i}$ represent the nondimensional speed of sound outside and inside of the jet, respectively. The relation between the two Mach numbers is determined using Crocco-Busemann's rule, i.e.
\begin{equation}
    M_{\mathrm{o}}=\dfrac{M_{\mathrm{i}}}{\sqrt{1+\dfrac{\gamma-1}{2}M_{\mathrm{i}}^2}}\nu^{1/2},
    \label{M_c}
\end{equation}
where $\nu$ denotes the temperature ratio between $\tilde{T}_{0,\mathrm{i}}$ and $\tilde{T}_{\mathrm{o}}$, and $\gamma$ represents the specific heat ratio.  
%\ccc{Description of the basic physical scenario.} 

Before proceeding to solve the problem, it is useful to examine the designed transfer function 
 using the Buckingham $\pi$ theorem. As mentioned above, eight physical variables are involved in this model, i.e. $\tilde{D}$, $\tilde{U}$, $\tilde{\rho}_{\mathrm{i}}$, $\tilde{\rho}_{\mathrm{o}}$, $\tilde{T}_{0,\mathrm{i}}$, $\tilde{T}_{\mathrm{o}}$, $\tilde{a}_\infty$, and the frequency of the acoustic wave $\tilde{\omega}$, while four independent dimensions can be found. Employing the Buckingham $\pi$ theorem, it is evident that the transfer function $H$ between the newly-excited instability wave and the acoustic waves is related to
four nondimensional parameters. These four parameters are chosen to be the Mach number $M_{\mathrm{i}}$, the Strouhal number $\omega$, the temperature ratio $\nu$, and the density ratio $\sigma$. The density ratio can be calculated using $\nu$ and $M_\mathrm{i}$ through the Crocco-Busemann's rule (\ref{M_c}). Therefore, there remain three independent nondimensional parameters and our objective is to find the transfer function $H\left(M_{\mathrm{i}},\omega,\nu\right)$ in the subsequent analysis.

%\ccc{The expression of the upstream forcing.} 
To have a realistic incident acoustic wave, we take the result obtained in the shock-instability interaction work~\citep{Myown, MyOwn_2}. Pack’s model was used to describe shock waves, while instability waves were calculated using the linear stability
analysis. The resulting nondimensionalized velocity potential function of the acoustic waves due to one-cell interaction can be obtained by performing the inverse Fourier transform, i.e.
\begin{equation}
    \phi_{\mathrm{in}}(x,y)=\frac{1}{2\pi}\int_{-\infty}^{\infty}D_{\mathrm{o}}(\lambda)\mathrm{exp}(-(\mathrm{i}\lambda x+\sqrt{\lambda^2-\omega^2 M_\mathrm{o}^2}y))\mathrm{d}\lambda.
    \label{initial_integraion}
\end{equation}
Herein $\lambda$ denotes the streamwise wavenumber and $D_{\mathrm{o}}(\lambda)$ represents a coefficient, which can be readily calculated when $M_{\mathrm{o}}$ and $\omega$ are provided. The detailed expression of $D_\mathrm{o}(\lambda)$ can be found in \citet{MyOwn_2}. The integration in (\ref{initial_integraion}) can be estimated (in the far field) by the saddle point method. The closed-form asymptotic potential function can be expressed as follows
  \begin{equation}
    \phi_{\mathrm{in}}(r,\theta)=\sqrt{\frac{{\omega M_{\mathrm{o}}}}{{2\pi}}}\frac{\mathrm{e}^{\mathrm{i}\omega(M_{\mathrm{o}}r-t-\pi/4)}}{\sqrt{r}}D_{\mathrm{o}}(-\omega M_{\mathrm{o}}{\rm c o s} \theta){\rm s i n} \theta ,
        \label{steepest_decent_way_p}
    \end{equation}
where $\theta$ is the observer angle with respect to the downstream direction and $r$ represents the distance from the screech source region to the nozzle lip. 

%\ccc{The source location.}
 As shown in (\ref{steepest_decent_way_p}), $\phi_{\mathrm{in}}(r,\theta)$ depends on $r$. Experimental observations indicated that the screech source was situated several shock cells downstream from the
nozzle exit~\citep{1994Suda,S.kaji,Sources}. Therefore, the streamwise distance between the source position to the nozzle lip is written as
\begin{equation}
    h=n s,
    \label{equ:n}
\end{equation}
where $n$ typically takes the value of 3, 4, or 5~\citep{1994Suda, X.D.Li,Sources}, and $s$ denotes the shock spacing, which can be calculated using Pack's model~\citep{1950Pack,Tam_shock_space}.

%\ccc{An asymptotic form of the potential function.}
 We consider the potential function of the upstream forcing $\Phi_{\mathrm{in}}(r,\theta)$ in the proximity of the nozzle lip, where the value of $y-1/2$ is relatively small compared to $|x|=h$. Therefore, the distance $r$ in (\ref{steepest_decent_way_p}) can be approximated as $r\approx h$. Similarly, we have $\sin \theta\approx(y-1/2)/h$ and $\cos\theta\approx1$. Consequently, in the vicinity of the nozzle, the potential function $\Phi_{\mathrm{in}}$ can be reformulated to a similar form to that in \citet{DGCrigh}, i.e.
\begin{equation}
    \Phi_{\mathrm{in}}(Y)=GY,
    \label{g_plus}
\end{equation}
where
\begin{equation}
    G=\sqrt{\frac{{\omega M_{\mathrm{o}}}}{{2\pi}h^{3}}}D_{\mathrm{o}}(\omega M_{\mathrm{o}}) \mathrm{e}^{\mathrm{i}\omega(M_{\mathrm{o}}h-\pi/4)},
    \label{equ:G}
\end{equation}
\begin{equation}
    Y=y-\frac{1}{2}.
\end{equation}

%\ccc{Basic assumptions before derivation.} 
The goal is therefore to find the response of the jet flow due to the upstream forcing $GY$ near the nozzle lip.  In this paper, we focus on the upper half-plane ($y>0$), while a similar analysis can be carried out for the lower half-plane.  Considering that the time-harmonic assumption is used, the factor $\mathrm{e}^{-\mathrm{i}\omega t}$ in all the time-dependent fields is omitted for clarity in the rest of this paper. Because we assume the initial base flow to be irrotational and inviscid both inside and outside of the vortex sheet~\citep{196batchelor}, the velocity potential induced by the incident acoustic wave near the nozzle lip can be described by a potential function. Therefore, taking into account (\ref{g_plus}), the total velocity potentials outside and inside the jet flow can be defined by
\begin{equation}
    \Phi=GY+\phi(x,y),
    \label{8}
\end{equation}
\begin{equation}
    \Psi=GY+\psi(x,y),
    \label{9}
\end{equation}   
 respectively, where $\phi$ and $\psi$ are the corresponding potential functions of the scattered sound fields in the region $y>1/2$ and $0\leq y\leq1/2$, respectively. 

%\ccc{Governing equations and boundary conditions.} 
It is known that $\phi$ and $\psi$ both satisfy the convective wave equation, i.e. 
\begin{equation}
    \mathbf{\nabla}^{2}\phi+\omega^2 M_{\mathrm{o}}^2\phi=0,
   \label{phi}
\end{equation}
\begin{equation}
    \mathbf{\nabla}^{2}\psi+\left( \omega+ \mathrm{i}M_{\mathrm{i}}\frac{\partial}{\partial x}\right)^2\psi=0.
    \label{psi}
\end{equation}
 Hard-wall conditions are imposed on $y=1/2$ for $-\infty<x \leq0$. Note that, strictly speaking, these conditions should be applied at $y=\tilde{D_0}/2\tilde{D}$, which can be either less than or greater than $1/2$ depending on whether the jet is underexpanded or overexpanded~\citep{tam_machwave}. However, as mentioned earlier, the difference between $\tilde{D_0}/2\tilde{D}$ and $1/2$ is relatively small compared to the distance $r$. To enable analysis progress, we neglect this difference and assume that the plate is nearly parallel to the boundary of the jet. Therefore, we obtain
\begin{equation}
    \frac{\partial \phi}{\partial y}=\frac{\partial \psi}{\partial y}=-G \quad (y=1/2,\ -\infty<x\leq0).
    \label{-G}
\end{equation}
The absence of decay in the forcing term $-G$ leads to convergence issues when the Fourier transform is applied to (2.13). To circumvent this problem,~\citet{DGCrigh} replaced (\ref{-G}) by
\begin{equation}
       \frac{\partial \phi}{\partial y}=\frac{\partial \psi}{\partial y}=-G\mathrm{e}^{\epsilon x},
       \label{-G2}
\end{equation}
where $\epsilon$ is a real and positive term, which retains the physical characteristics of the upstream forcing near the nozzle lip but removes the forcing further upstream. We follow the same treatment here. 

The kinematic and dynamic boundary conditions can be used on the vortex sheet at $y=1/2, 0<x<\infty$, i.e.
\begin{equation}
    \left(-\mathrm{i} \omega +\frac{\partial}{\partial x}\right)\frac{\partial\phi}{\partial y}=-\mathrm{i}\omega \frac{\partial \psi}{\partial y},
    \label{displacement}
\end{equation}
\begin{equation}
    -\mathrm{i}\omega\sigma\phi= \left(-\mathrm{i} \omega +\frac{\partial}{\partial x}\right)\psi.
    \label{pressure}
\end{equation}
%\ccc{Fourier transformation and the solution to the scattered field.} 
Fourier transforms $\phi(k,y), \psi(k,y)$ are defined by
    \begin{gather}
	\phi(k,y)=\int_{-\infty}^{+\infty} \phi(x,y)\mathrm{e}^{\mathrm{i} k x}\mathrm{d} x,\\
    \psi(k,y)=\int_{-\infty}^{+\infty} \psi(x,y)\mathrm{e}^{\mathrm{i} k x}\mathrm{d} x,
	\label{equ:fourier}
    \end{gather}
where $k$ is the wavenumber in the streamwise direction. Multiplying~(\ref{phi}) and~(\ref{psi}) with $\mathrm{e}^{\mathrm{i}kx}$ and subsequently integrating both expressions with respect to the variable $x$, solutions to (\ref{phi}) and (\ref{psi}) can be found upon invoking the far-field boundary condition, i.e.
\begin{equation}
    \phi(k,y)=\phi^{*}(k)\mathrm{e}^{-\gamma_1(k)y},
    \label{equ:20}
\end{equation} 
\begin{equation}
    \psi(k,y)=\psi_{\mathrm{s}}^{*}(k)\sinh\gamma_2(k)y+\psi_\mathrm{c}^{*}(k)\cosh\gamma_2(k)y,
\end{equation}
where
\begin{equation}
  \gamma_2(k)=\sqrt{k^2-(\omega+k)^2M_{\mathrm{i}}^2} \quad  \mathrm{and} \quad \gamma_1(k)=\sqrt{k^2-\omega^2 M_{\mathrm{o}}^2}.
\end{equation}
Herein, $\phi^*(k)$, $\psi_{\mathrm{s}}^*(k)$, and $\psi_{\mathrm{c}}^{*}(k)$ are functions of $k$, whose detailed forms will be specified subsequently.
To satisfy the far-field condition, the real part of $\gamma_{1}$ should be positive, and its branch cut will be specified later.
It is well-known that both symmetric (varicose) and antisymmetric (sinuous) instability modes 
can exist inside the jet. But for jet flows from high-aspect-ratio rectangular nozzles, the sinuous mode is dominant~\citep{2019EDINGTON}. Therefore, we consider the antisymmetric mode in this study, while the symmetric mode can be examined in a similar way should interest arise.
In this case, $\psi(k,y)$ reduces to
\begin{equation}
    \psi(k,y)=\psi^{*}_\mathrm{s}(k)\sinh\gamma_2(k)y.
    \label{equ:24}
\end{equation}

%\ccc{Derivation of the Wiener-Hopf Equation.} 
In what follows, we use the Wiener-Hopf method to solve this boundary-value problem specified by (\ref{equ:20}), (\ref{equ:24}) and boundary conditions (\ref{-G2}), (\ref{displacement}), and (\ref{pressure}). Using (\ref{-G2}), we can obtain
\begin{equation}
    \phi^{\prime}(k,1/2)=\phi_{+}^{\prime}(k,1/2)+\frac{\mathrm{i}G}{k-\mathrm{i}\epsilon},
    \label{combine3}
\end{equation}
 where the prime represents the first derivative with respect to $y$ and $\phi_{+}(k,1/2)$ is defined by
 \begin{equation}
    \phi_{+}(k,1/2)=\int_{0}^{\infty}\phi(x,1/2)\mathrm{e}^{\mathrm{i}k x}\mathrm{d}x.
 \end{equation}
Considering the kinematic boundary condition across the vortex sheet, i.e. (\ref{displacement}), and the boundary conditions imposed on the nozzle lip, i.e. (\ref{-G2}),  we have
\begin{equation}
    (k-\mathrm{i}\epsilon)\psi^{*}_\mathrm{s}(k)\cosh\gamma_2(k)/2-\frac{G}{\omega}\epsilon=-\left(1+\frac{k}{\omega}\right)(k-\mathrm{i}\epsilon)\frac{\gamma_1(k)}{\gamma_2(k)}\phi^{*}(k)\mathrm{e}^{-\gamma_1(k)/2}.
    \label{combine2}
\end{equation}
Following \citet{DGCrigh}, the term $\dfrac{G}{\omega}\epsilon$ on the left-hand side of (\ref{combine2}) is omitted, considering that $\epsilon$ is a small term.

%\ccc{One crucial step of the Wiener-Hopf technique.}
 Now we define a new function related to the difference between the pressures outside and inside the vortex sheet as $D(k,1/2)$, i.e.
\begin{equation}
    D(k,1/2)=\int_{-\infty}^{\infty} \left(-\mathrm{i}\omega \sigma\phi(x,1/2)-\left(-\mathrm{i}\omega+\frac{\partial}{\partial x}\right)\psi(x,1/2)\right)\mathrm{e}^{\mathrm{i}k x}\mathrm{d}x.
\end{equation}
Due to the continuity of the pressure across the vortex sheet, i.e. (\ref{pressure}), the pressure difference vanishes when $x>0$. Therefore, $D(k,1/2)$ reduces to
\begin{equation}
    \begin{aligned}
          D(k,1/2)
    &= -\mathrm{i}\omega \sigma\phi_{-}(k,1/2)-(-\mathrm{i}\omega-\mathrm{i}k)\psi_{-}(k,1/2)+\psi_0,
    \label{combine4}
    \end{aligned}
\end{equation}
where
\begin{gather}
    \phi_{-}(k,1/2)=\int_{-\infty}^{0}\phi(x,1/2)\mathrm{e}^{\mathrm{i}k x}\mathrm{d}x \quad \mathrm{and} \quad \psi_{-}(k,1/2)=\int_{-\infty}^{0}\psi(x,1/2)\mathrm{e}^{\mathrm{i}k x}\mathrm{d}x,
\end{gather}
and $\psi_0$ denotes the finite value of $\psi(x,y)$ at the position $(0,1/2)$.
Substituting (\ref{equ:20}) and (\ref{equ:24}) into (\ref{combine4}), we obtain
\begin{equation}
    D(k,1/2)=-\mathrm{i}\omega \sigma \phi^{*}(k)\mathrm{e}^{-\gamma_1(k)/2}-(-\mathrm{i}\omega-\mathrm{i}k)\psi^{*}_\mathrm{s}(k)\sinh\gamma_2(k)/2.
    \label{combine1}
\end{equation}

%\ccc{Wiener-Hopf equation, the kernel function, and usage of Liouville's theorem.} 
Combining~(\ref{combine3}), (\ref{combine2}), (\ref{combine4}), and (\ref{combine1}) yields
\begin{equation}
    \phi_{+}^{\prime}(k,1/2)+\frac{\mathrm{i}G}{k-\mathrm{i}\epsilon}=K(k)F_{-}(k),
    \label{Wiener-Hopf_equation}
\end{equation}
where
\begin{equation}
     K(k)=\frac{1}{\dfrac{\sigma}{\gamma_1(k)}+\dfrac{(1+k/\omega)^2\tanh\gamma_2(k)/2}{\gamma_2(k)}},
\end{equation}
\begin{equation}
    F_{-}(k)=(1+k/{\omega})\psi_{-}(k,1/2)-\sigma \phi_{-}(k,1/2)-{\mathrm{i}}\psi_0/{\omega}.
    \label{equ:36}
\end{equation}   
%Here $\sigma=M_{\mathrm{o}}^2/M_{\mathrm{i}}^2$. 
Equation~(\ref{Wiener-Hopf_equation}) represents the Wiener-Hopf equation we aim to solve, with the kernel given by $K(k)$. $1/K(k)=0$ represents the dispersion relation of the antisymmetric mode of the jet instability wave. It can be seen that as $|k|\rightarrow\infty$, 
$|K(k)|\rightarrow|k|^{-1}$. Now, following the procedures of the Wiener-Hopf method, we assume that the kernel can be decomposed into
\begin{equation}
    K(k)=K_{+}(k)K_{-}(k),
\end{equation}
where $K_{+}(k)$ and $K_{-}(k)$ are analytic and non-zero in the upper and lower half-planes of $k$, respectively. Meanwhile, both $K_{+}(k)$ and $K_{-}(k)$ behave as $O(k^{-1/2})$ as $k$ approaches infinity. The actual decomposition will be shown in section \ref{sec:B}. Routine use of Wiener-Hopf decomposition leads to
\begin{equation}
    \frac{\phi_{+}^{\prime}(k,1/2)}{K_{+}(k)}+\left(\frac{1}{K_{+}(k)}-\frac{1}{K_{+}(\mathrm{i}\epsilon)}\right)\frac{\mathrm{i}G}{k-\mathrm{i}\epsilon}=E(k),
    \label{equ:wh_plus}
\end{equation}
\begin{equation}
    K_{-}(k)F_{-}(k)-\frac{1}{K_{+}(\mathrm{i}\epsilon)}\frac{\mathrm{i}G}{k-\mathrm{i}\epsilon}=E(k).
\end{equation}
Here $E(k)$ is an entire function of $k$. Due to the Kutta condition~\citep{Crighton_Kutta}, $\phi(x,1/2)$ behaves like $x^{3/2}$ near the nozzle lip. The corresponding value of $ \phi^\prime(x,1/2)$ is therefore $O(x^{1/2})$. It follows that $\phi_{+}^{\prime}(k,1/2)\rightarrow k^{-3/2}$ as $k\rightarrow\infty$~\citep{erfelyi}. Thus, from Liouville's theorem,
\begin{equation}
    E(k)\equiv 0.
\end{equation}

%\ccc{The singularity.} 
From (\ref{combine3}) and (\ref{equ:wh_plus}), $\phi^{\prime}(k,1/2)$ can be readily solved, i.e.
\begin{equation}
    \phi^{\prime}(k,1/2)=\frac{\mathrm{i}G}{k-\mathrm{i}\epsilon}\frac{K_{+}(k)}{K_{+}(\mathrm{i}\epsilon)}.
    \label{equ:phi_plus}
\end{equation}
It can be seen that $\phi^{\prime}(k,1/2)$ is $O(k^{-3/2})$ as $|k|$ approaches infinity in the upper half plane. The corresponding velocity component ${\phi^\prime(x,y)}$ behaves as $x^{1/2}$ when $x\rightarrow 0$, which leads to singularity in ${\psi^\prime(x,y)}$ due to (\ref{displacement}). Therefore,
 (\ref{equ:phi_plus}) leads to singular behaviour when $|k|\rightarrow \infty$, where (2.37) can be approximated by \citep{DGCrigh}
\begin{equation}
    \frac{\mathrm{i}G}{K_{+}(\mathrm{i}\epsilon)}\frac{K_{+}(k)}{k}.
    \label{kutta_1}
\end{equation}
This part will be further discussed in section \ref{sec:2D}, where the unsteady Kutta condition
is used to determine the new instability wave by removing the emerging singularity.

\subsection{The kernel decomposition}
\label{sec:B}
\begin{figure}
    \centering
    \includegraphics[width = 0.9\textwidth]{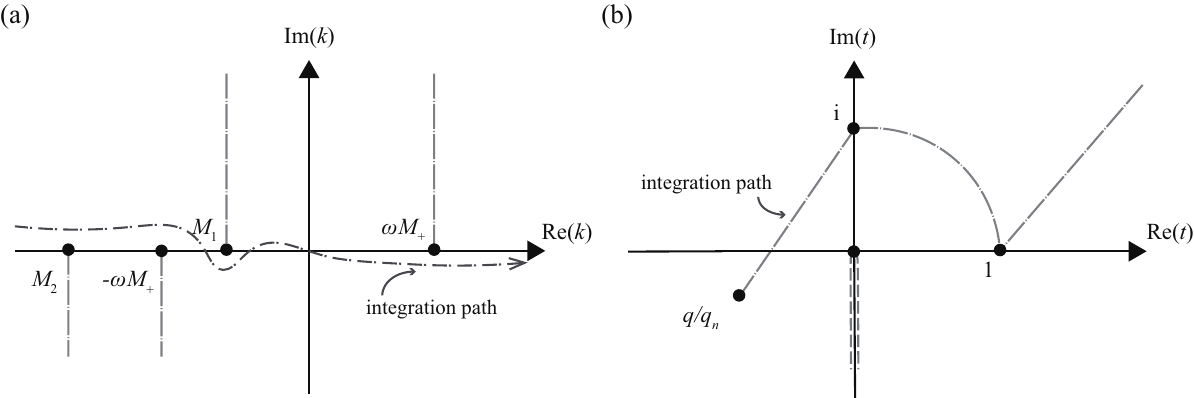}
    \caption{(a) Schematic of the branch points, branch cuts, and the integration path in the complex $k$ plane. (b) Schematic of the deformed integration path in the complex $t$ plane.}
    \label{fig:branch}
\end{figure}

%\ccc{Define the branch cuts.}  
In this section, we factor the kernel $K(k)$ into $K_{+}(k)$ and $K_{-}(k)$, which is the essential step of the Wiener-Hopf technique. Before proceeding with the decomposition of the kernel, it is necessary to properly define the branch cuts of $K(k)$. In order to ensure that the real part of $\gamma_{1}$ is positive as $|k|\rightarrow \infty$ along the integration path (which becomes relevant when conducting the inverse Fourier transform if we are interested in calculating the scattered sound field),  the branch
cuts of  $\gamma_{1}$  passing the branch points $k=\pm \omega M_{\mathrm{o}}$ are chosen to extend to the upper and lower half-plane, respectively, as shown in figure \ref{fig:branch}(a). The branch is selected in such a way that $\mathrm{arg}(\gamma_1)=-\pi/2$ at $k=\mathrm{i}$.
%In this way, $\mathrm{arg}(\gamma_{+})\rightarrow0$ as $|\lambda|\rightarrow \infty$. 
On the other hand, the branch points of $\gamma_{2}$ are determined as $M_{1}=-\omega M_{\mathrm{i}}/(M_{\mathrm{i}}+1)$ and $M_{2}=-\omega M_{\mathrm{i}}/(M_{\mathrm{i}}-1)$. The branch cuts passing $M_{1}$ and $M_{2}$ are chosen to extend to the lower and upper
half-plane, respectively, for the convenience of the kernel decomposition. The branch of $\gamma_2$ is selected to ensure that $\mathrm{arg}(\gamma_2)=-\pi/2$ at $k=\mathrm{i}$. Due to the choice of the branch cuts, we can find that $\gamma_2(k)/\gamma_1(k)\rightarrow \sqrt{1-M_{\mathrm{i}}^2}$, as $\omega\rightarrow 0$ and $k\rightarrow O(1)$.

%\ccc{Decomposition to $\gamma_2(k)$.} 
The kernel can be rewritten as
\begin{equation}
    K(k)=\frac{\gamma_2(k)}{\sigma \gamma_2(k)/\gamma_1(k)+(1+k/\omega)^2\tanh\gamma_2(k)/2}.
\end{equation}
A proper decomposition of $\gamma_2(k)$ is evident, i.e. $\gamma_2(k)=\sqrt{k-M_1}\sqrt{k-M_2}$. Thus, in what follows we only focus on the denominator, which we refer to as $Q(k)$. It can be seen that $Q(k)$ depends on the dispersion relation of the antisymmetric mode of the jet instability wave.
This dispersion relation describes several types of waves present in the jet, determined by its structural characteristics such as branches, zeros, and poles. These information can be maintained if the kernel
is decomposed by means of matched asymptotic expansions~\citep{DGCrigh, Crighton_2001}. To use this strategy, a small parameter should be specified as the basis for the expansion. This parameter should not only be small but can also be
contained by the kernel. We introduce $S=\omega(M_{\mathrm{i}}-1)^2$ as this parameter. In what follows, the complicated kernel is replaced by a simpler one by asymptotic expansions based on $S$. Overlapping approximations are subsequently applied to guarantee the validity of the decomposition over the whole range of the wavenumber $k$.
The final result is obtained by a multiplicative composite approximation.

%\ccc{The asymptotic form of $Q(k)$ under one assumption. } For $Q(k)$,
 We first assume $k=O(1)$ and let $\omega\rightarrow 0$. The first term of $Q(k)$, i.e. $\sigma\gamma_2(k)/\gamma_1(k)$ reduces to $\sigma \sqrt{1-M_{\mathrm{i}}^2}$ and $Q(k)$ can be approximated as 
\begin{equation}
    \begin{aligned}
          Q(k)%&\approx\sigma\sqrt{1-M_{\mathrm{i}}^2}+\left(1+\frac{k}{\omega}\right)^2\tanh(\sqrt{(1-M_{\mathrm{i}}^2)k^2}/2) \\ 
    &\approx \left(\frac{k}{\omega}\right)^2\tanh\sqrt{(1-M_{\mathrm{i}}^2)k^2}/2.
    \label{appr1}
    \end{aligned}
\end{equation}
The term $\sqrt{(1-M_{\mathrm{i}}^2)k^2}$ is supposed to maintain the branch cut chosen for $\gamma_2(k)$. For brevity, we rewrite $\sqrt{(1-M_{\mathrm{i}}^2)k^2}$ as $\sqrt{1-M_{\mathrm{i}}^2}k^{*}$, where the
script $^*$ represents that this parameter retains the specified branch cuts.

%\ccc{The asymptotic form of $Q(k)$ under one other assuption.} 
Now, we introduce the small parameter $S=\omega(M_{\mathrm{i}}-1)^2$ and a new scaled wavenumber according to
\begin{equation}
    q=k/S^{\frac{3}{4}}.
\end{equation}
By assuming $S\rightarrow 0$ and $q\rightarrow O(1)$, we can expand $Q(k)$ in (2.39) based on the small parameter $S$ (using methods of Taylor expansion), i.e.
\begin{equation}
\begin{aligned}
    Q(k)&\approx\sqrt{1-M_{\mathrm{i}}^2}\bigg(\sigma-\frac{\sigma M_{\mathrm{i}}^2S^{1/4}}{(M_{\mathrm{i}}-1)^2(1-M_{\mathrm{i}}^2)q^{*}}+\frac{1}{2}(M_{\mathrm{i}}-1)^4S^{1/4}q^{2}q^{*}\bigg),
      \label{appr2}
\end{aligned}
\end{equation}
where $q^*=k^*/S^{\frac{3}{4}}$. 
Note that despite the small parameter $S\sim o(1)$, the quantities $S^{1/4}\sim O(1)$. To put this into perspective, let us consider an example with $M_{\mathrm{i}}=1.1$. At the screech frequency, employing Powell's original model yields $S=0.028$. However, $S^{1/4}=0.41$, which is of the order of unity. The scale separation can be shown more clearly as $M_{\mathrm{i}}$ increases. This demonstrates that although $S$ itself may be small, its corresponding fractional power can still have magnitudes of the order of unity in our interested application. 

%\ccc{Overlapping.} 
We can show that (\ref{appr1}) and (\ref{appr2}) do overlap, as $k\rightarrow 0$ and $q\rightarrow \infty$, respectively, with common value $\dfrac{1}{2}\sqrt{1-M_{\mathrm{i}}^2}(M_{\mathrm{i}}-1)^4S^{1/4}q^{2}q^{*}$. A multiplicative composite approximation then reads
\begin{equation}
\begin{aligned}
    Q(k)&\approx\sqrt{1-M_{\mathrm{i}}^2} \frac{\tanh \dfrac{1}{2}\sqrt{1-M_\mathrm{i}^2} k}{\dfrac{1}{2}\sqrt{1-M_\mathrm{i}^2} k} \frac{1}{q^{*}}R(q),
      \label{decom_2}
\end{aligned}
\end{equation}
where
\begin{equation}
    R(q)=\sigma q^{*}-\frac{\sigma M_{\mathrm{i}}^2S^{1/4}}{(M_{\mathrm{i}}-1)^2(1-M_{\mathrm{i}}^2)}+\frac{1}{2}(M_{\mathrm{i}}-1)^4S^{1/4}q^{4}.
    \label{2.46}
\end{equation}

\begin{figure}
    \centering
    \includegraphics[width = 0.5\textwidth]{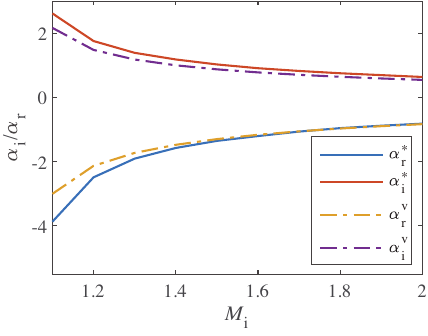}
    \caption{Comparison between the wavenumbers of the sinuous mode of the jet instability wave obtained by the decomposed kernel (\ref{decom_2}) and the dispersion relation. $\alpha^{*}$ and $\alpha^{v}$ denote the wavenumber calculated from the decomposed kernel and $M(k)$, respectively, while the subscript $\rm_r$ and $\rm_i$ represent the real and imaginary part of the parameter, respectively. $\sigma$ is assumed to be $1$.}
    \label{fig:decom}
\end{figure}

%\ccc{Demonstration to the validity of the decomposition.} 
To validate the decomposition, the wavenumber obtained by letting $Q(k)$ in (\ref{decom_2}) equal to 0, which corresponds to the wavenumber of the sinuous mode of the jet instability wave, is compared to that numerically solved by the dispersion relation $1/K(k)=0$. As shown in figure~\ref{fig:decom}, the agreement between the two wavenumbers is good when $M_{\mathrm{i}}>1.2$. Therefore, in the subsequent analysis, our focus will be in the range of $M_{\mathrm{i}}>1.2$.

%\ccc{Decomposition of two functions contained in the kernel function.} 
We are now in a position to factor the function $Q(k)$. From~\citet{Noble}, $\dfrac{\tanh \sqrt{1-M_\mathrm{i}^2} k/2}{\sqrt{1-M_\mathrm{i}^2} k/2}$  can be decomposed to
\begin{equation}
    \begin{aligned}
        \frac{\tanh \dfrac{1}{2}\sqrt{1-M_\mathrm{i}^2} k}{\dfrac{1}{2}\sqrt{1-M_\mathrm{i}^2} k} 
    &= T_{+}(k)T_{-}(k),
    \end{aligned}
\end{equation}
where
\begin{gather}
    T_+(k)=\frac{\Gamma(1/2+\sqrt{M_\mathrm{i}^2-1}k/2\pi)}{\pi^{1/2}\Gamma(1+\sqrt{M_\mathrm{i}^2-1}k/2\pi)},\quad T_-(k)=\frac{\Gamma(1/2-\sqrt{M_\mathrm{i}^2-1}k/2\pi)}{\pi^{1/2}\Gamma(1-\sqrt{M_\mathrm{i}^2-1}k/2\pi)}.
     \label{tank/k}
\end{gather}
Here $\Gamma$ represents the Gamma function. The decomposition of $1/q^*$ reads
\begin{equation}
    \begin{aligned}
         \frac{1}{q^*} 
    &=\mathcal{Q}_{+}(q)\mathcal{Q}_{-}(q),
    \end{aligned}
\end{equation}
where
\begin{equation}
    \mathcal{Q}_{+}(q)=\frac{1}{\sqrt{q+0}},\quad \mathcal{Q}_{-}(q)=\frac{1}{\sqrt{q-0}}.
    \label{q_star}
\end{equation}
Naturally, the branch cut of $\mathcal{Q}_{+}$ and $\mathcal{Q}_{-}$ is from $-0$ to $-\mathrm{i}\infty$ and from $+0$ to $+\mathrm{i}\infty$, respectively. 

%\ccc{Decompose $R(q)$. This paragraph is the first step.} 
The decomposition of $R(q)$, on the other hand, is very difficult due to the presence of $q^{*}$ in (\ref{2.46}).
% However, considering that for the polynomial $P(q)$ defined by
%\begin{equation}
 %   P(q)=R(q)-\sigma q^{*}=-\frac{\sigma M_{\mathrm{i}}^2S^{1/4}}{(M_{\mathrm{i}}-1)^2(1-M_{\mathrm{i}}^2)}+\frac{1}{2}(M_{\mathrm{i}}-1)^4S^{1/4}q^{4},
%\end{equation}
%zeros of $P(q)^2-\sigma^2q^2$ include all the zeros of $R(q)$. Therefore, we may try to decompose $R(q)$ by increasing its order.
To decompose $R(q)$, we let $q$ be real first~\citep{DGCrigh}, so that $q^{*}=|q|$. This is reasonable considering that the overlap of the two half-planes will converge to the real axis in the complex $k$ plane.
Assuming that $R(q)$ can be decomposed to $R(q)=R_{+}(q)R_{-}(q)$, and taking the first derivative of $\mathrm{ln}R(q)$, we obtain
\begin{equation}
    \frac{\mathrm{d}}{\mathrm{d}q}\mathrm{ln}R_{+}(q)+\frac{\mathrm{d}}{\mathrm{d}q}\mathrm{ln}R_{-}(q)=\frac{4q^3+a_1 |q|/q}{q^4+a_1|q|+a_0},
    \label{ln form_2}
\end{equation}
where
\begin{gather*}
    a_0=2\sigma M_{\mathrm{i}}^2(M_{\mathrm{i}}-1)^{-6}(M_{\mathrm{i}}^2-1)^{-1},\\
    a_1=2\sigma (M_{\mathrm{i}}-1)^{-4}S^{-1/4}.
\end{gather*}
Multiplying the numerator and the denominator of the right-hand side term in (\ref{ln form_2}) by $q^4-a_1|q|+a_0$, equation (\ref{ln form_2}) can be written as
\begin{equation}
    \frac{\mathrm{d}}{\mathrm{d}q}\mathrm{ln}R_{+}(q)+\frac{\mathrm{d}}{\mathrm{d}q}\mathrm{ln}R_{-}(q)=\frac{(4q^3+a_1 |q|/q)(q^4-a_1|q|+a_0)}{(q^4+a_0)^2-a_1^2q^2}.
    \label{ln form}
\end{equation}

%\ccc{Use Ferrari's formula to reorganize (2.54).} 
The zeros of the denominator $(q^4+a_0)^2-a_1^2q^2$ are defined as $q_n, n=1,...,8$, which can be obtained by Ferrari's formula. The details are provided in Appendix A.
With the knowledge of these zeros, we can now reorganize~(\ref{ln form}) as
\begin{equation}
    \begin{aligned}
         \frac{\mathrm{d}}{\mathrm{d} q} \ln \left(\frac{R_{+}(q)}{\prod_{n=5}^8\left(q-q_n\right)^{1/2}}\right)+\frac{\mathrm{d}}{\mathrm{d} q} \ln& \left(\frac{R_{-}(q)}{\prod_{n=1}^4\left(q-q_n\right)^{1/2}}\right) \\
&=\sum_{n=1}^4 \xi_n  \frac{|q|}{\left(q-q_n\right)}+
\sum_{n=5}^8\xi_n \frac{|q|}{\left(q-q_n\right)}+\frac{a_1 }{a_0 }\frac{|q|}{q},
\label{R+R-}
    \end{aligned}
\end{equation}
where
\begin{equation}
    \xi_n =\frac{-3a_1q_n^4+a_0a_1}{2q_n^2(4q_n^6+4a_0q_n^2-a_1^2)}.
\end{equation}

The detailed derivation from (\ref{ln form}) to (\ref{R+R-}) can be found in Appendix B. For brevity, we can define $\xi_9=a_1/a_0$ and $q_9=0$. With these definitions, it is straightforward to find 
\begin{equation}
    \sum_{n=1}^9 \xi_n q_n=\sum_{n=1}^9 \xi_n=0.
    \label{equ:relation1}
\end{equation}
%\ccc{Obtain $R_{+}(q)$.} 
An addictive decomposition of $|q|$ is now introduced, i.e.
\begin{eqnarray}
    |q|&=&Q_{+}(q)+Q_{-}(q), 
    \label{equ:addictive decomposition}
\end{eqnarray}
where
\begin{equation}
    Q_{+}(q)=q/2+\frac{\mathrm{i}q}{\pi}\ln_{+}q,\quad Q_{-}(q)=q/2-\frac{\mathrm{i}q}{\pi}\ln_{-}q.
\end{equation}
The branch cuts of $\ln_{+}$ and $\ln_{-}$ extend from $0$ to $-\mathrm{i}\infty$ and from $0$ to $+\mathrm{i}\infty$, respectively. The choice of the branch cut leads to
\begin{equation}
    Q_+(q)=Q_{-}(-q).
    \label{equ:relation2}
\end{equation}
Using the addictive decomposition (\ref{equ:addictive decomposition}) and the two identities (\ref{equ:relation1}) and (\ref{equ:relation2}),
the right-hand side of (\ref{R+R-}) can obtain a similar addictive factorization, and $R_{+}(q)$ can be finally found as
%\begin{equation}
 %   R_{+}(q)=C_{+}\prod_{n=1}^4(q-q_n)^{1/2}\mathrm{exp}\left(\frac{\mathrm
  %  i}{\pi}\Sigma_{n=1}^{4}q_n^2\int_{-\infty}^{q}\frac{(\beta_n+\gamma_n/q)\ln_{+}(q/q_n)}{q^2-|q_n||q_{n+4}|-(\mathrm{Re}(q_n)+\mathrm{Re}(q_{n+4}))q}\mathrm{d}q \right),
%\end{equation}
%\begin{equation}
%    R_{+}(q)=\prod_{n=1}^4(q-q_n)^{1/2}\mathrm{exp}\left(\frac{\mathrm
%    i}{\pi}\Sigma_{n=1}^{4}q_n^2\int_{-\infty}^{q}\frac{(\beta_n+\gamma_n/q)\ln_{+}(q/q_n)}{(q-q_n)(q-q_{n+4})}\mathrm{d}q \right),
%\end{equation}
%where 
%\begin{gather}
%    \beta_n=\alpha_n+\alpha_{n+4},\\
%    \gamma_n=-(\alpha_n q_{n+4}+\alpha_{n+4}q_n).
%\end{gather}
\begin{equation}
    R_{+}(q)=\prod_{n=5}^8(q-q_n)^{1/2}\mathrm{exp}\left(-\frac{\mathrm
    i}{\pi}\sum_{n=1}^{8}\xi_n q_n\int_{q/q_n }^{\infty}\frac{\ln_{+}t}{t-1}\mathrm{d}t +\sum_{n=5}^{8}{\xi_n q_n}\int_{q/q_n }^{\infty}\frac{1}{t-1}\mathrm{d}t \right).
    \label{Rplus}
\end{equation}
The detailed derivation from (\ref{R+R-}) to (\ref{Rplus}) can be found in Appendix B.
%\ccc{Evaluate $R_{+}(q)$ by deforming the integral path.} 

To evaluate the integral in (\ref{Rplus}), as shown in figure \ref{fig:branch}(b), we deform the integration path in (\ref{Rplus}) from $q/q_n$ to the point $\mathrm{e}^{\pi\mathrm{i}/2}$, then along a unit circle to reach $q/q_n=1$ situated on the real axis, and finally along a ray from $q/q_n=1$ to $\infty$. Using the identity (\ref{equ:relation1}), we can evaluate (\ref{Rplus}) to obtain
%Note that (\citet{table} p. 532)
%\begin{equation}
 %   \int_{1}^{\infty}\frac{\ln_{+}(t)}{(t^2-t)}\mathrm{d}t=\frac{\pi^2}{6},
  %  \label{relation1}
%\end{equation}
%and
%\begin{equation}
 %  \int_{0}^{\pi/2}\frac{x}{\mathrm{e}^{\mathrm{i}x}-1}\mathrm{d}x=-\frac{1}{16}\pi^2-(\mathrm{C}+\frac{\pi}{4}\ln2)\mathrm{i},
 %  \label{relation 2}
 %\end{equation}
%where $\mathrm{C}$ is Catalan's constant~\citep{Catalan}. Using the identities shown in (\ref{relation1}) and (\ref{relation 2}) when deforming the integration path, we can evaluate (\ref{Rplus}) to obtain
\begin{equation}
\begin{aligned}
       R_{+}(q)=\prod_{n=5}^8(q-q_n)^{1/2}\mathrm{exp}\bigg(&-\frac{\mathrm
    i}{\pi}\sum_{n=1}^{8}\xi_n q_n \mathrm{Li}_2(1-q/q_n) -
    \sum_{n=5}^{8}{\xi_n q_n}\mathrm{ln}_+ \frac{q}{q_n}\bigg).
    \label{Rplus2}
\end{aligned}   
\end{equation}
%where $\Xi=\sum_{n=1}^{8} \xi_n q_n$ and the new function $\mathcal{L}$ is defined by %A further investigation then leads to $\beta\equiv 0$.
%\begin{equation}
   % \mathcal{L}(q)=-\mathrm{Li}_2(1-q)-\frac{1}{2}\mathrm{ln}_{+}^2(q).
%\end{equation}
Here $\mathrm{Li}_2$ represents the polylogarithm function of the second order~\citep{table}. %Therefore, the function $R_{+}(q)$ can be written as
%\begin{equation}
 %   \begin{aligned}
  %         R_{+}(q)=\prod_{n=5}^8(q-q_n)^{1/2}&\mathrm{exp}\bigg(-\frac{\mathrm
 %       i}{\pi}\sum_{n=1}^{8}\xi_n q_n(\mathcal{L}(\mathrm{e}^{\pi\mathrm{i}/2})-\mathcal{L}(q/q_n)) \bigg),
 %       \label{Rplus2_2}
 %   \end{aligned}   
 %   \end{equation}

%\ccc{Expression of $K_{+}(k)$.}
 Using  (\ref{tank/k}), (\ref{q_star}), and (\ref{Rplus2}) we obtain the final form of $K_{+}$, i.e.
\begin{equation}
    K_{+}(k)=\frac{\sqrt{k-M_2}}{T_{+}(k)R_{+}(k/S^{3/4})\mathcal{Q}_{+}(k/S^{3/4})}.
    \label{kernel_K}
\end{equation}
A complete expression of $K_{+}$ is ,
\begin{equation}
    \begin{aligned}
        K_{+}(k)=\sqrt{\pi}&\frac{\Gamma(1+\sqrt{M_\mathrm{i}^2-1}k/2\pi)}{\Gamma(1/2+\sqrt{M_\mathrm{i}^2-1}k/2\pi)}\frac{\sqrt{k-M_2}\sqrt{k/S^{3/4}+0}}{\prod_{n=5}^8(q-q_n)^{1/2}}
        \\
            &\mathrm{exp}\bigg(\frac{\mathrm
            i}{\pi}\sum_{n=1}^{8}\xi_n q_n \mathrm{Li}_2(1-q/q_n) +
            \sum_{n=5}^{8}{\xi_n q_n}\mathrm{ln}_+ \frac{q}{q_n}\bigg).
    \label{kernel_K_2}  
    \end{aligned} 
\end{equation}
As $k\rightarrow 0$, the Taylor expansion of $K_{+}(k)$ reduces to
\begin{equation}
    K_{+}(k)\approx\frac{\mathrm{i}\pi\sqrt{M_2}}{\mathcal{C}S^{3/8}}{\sqrt{k}}+O(k^{3/2}).
    \label{K_star_near_zero}
\end{equation}
 Here, the constant $\mathcal{C}$ is defined as $\mathcal{C}=\prod_{n=5}^{8}q_n^{1/2}$. Equation (\ref{K_star_near_zero}) will be used to determine $\epsilon$ in section \ref{sec:2D}.

%After some simple algebras, it can be proved that $\beta=0$. 

 \subsection{Instability waves excited by the acoustic waves at the nozzle lip}
 \label{sec:2D}
%\ccc{Determine the form of the newly-excited instability waves.} 
Under the excitation of the acoustic wave, new instability waves would emerge near the nozzle lip~\citep{Crighton_Kutta}. We can determine this instability wave by imposing the unsteady Kutta condition at the nozzle lip~\citep{Crighton_asabrunch}. Following~\citet{Crighton_1972} and considering that we focus on the upper half-plane ($y>1/2$), we assume that the excited instability wave can be represented by
\begin{equation}
    \phi^{*}=\mathcal{A}\mathrm{e}^{-\mathrm{i}\mathcal{K}x-\gamma(\mathcal{K})y}, 
    \label{equ:instability}
\end{equation}
where $\mathcal{A}$ represents the amplitude of the instability wave and $\gamma(\mathcal{K})$ is defined by $\gamma(\mathcal{K})=\sqrt{\mathcal{K}^{2}-\omega^2 M_{\mathrm{o}}^2}$. The parameter $\mathcal{K}$ denotes the wavenumber of the instability waves of the antisymmetric mode, and its determination is outlined in Appendix A. The choice of the form in (\ref{equ:instability}) is believed to be reasonable, particularly considering the sinuous nature exhibited by both the upstream forcing and the scattered acoustic field.
 
%\ccc{Outlines of the methods to obtain $A_1$.} 
The instability wave then propagates downstream and subsequently interacts with the shock structures \citep{1952Powell,1994Suda,2021Edington}. This interaction leads to the emission of sound, which serves as the upstream forcing and completes the screech feedback loop.
To determine the instability amplitude $\mathcal{A}$ in (\ref{equ:instability}), we first calculate the scattered field resulting from the newly-excited instability wave at the nozzle lip. The singular component within the resulting scattered field needs to be eliminated using the singularity in (\ref{kutta_1}). This process fixes the value of $\mathcal{A}$ in order to satisfy the unsteady Kutta condition.

% \ccc{Repeating the derivation in section 2.1 to find the singularity of the scattered field due to the instability wave.} 
Replacing the forcing term $GY$ in (\ref{8}) and (\ref{9}) with $\phi^{*}$ while following the same procedures outlined in section~\ref{subsection:base assumptions}, we obtain the scattered field due to the presence of the instability wave, i.e.
  \begin{equation}
      \phi^{*^\prime}_{+}(k,1/2)=\frac{\mathrm{i}\mathcal{A}\gamma(\mathcal{K})\mathrm{e}^{-\gamma(\mathcal{K})/2}}{k-\mathcal{K}}\left(1-\frac{K_{+}(k)}{K_{+}(\mathcal{K})}\right),
  \end{equation}
The component of $\phi^{*^\prime}_{+}(k)$ similar to (\ref{kutta_1}) that leads to singularity is
\begin{equation}
    -\frac{\mathrm{i}\mathcal{A}\gamma(\mathcal{K})\mathrm{e}^{-\gamma(\mathcal{K})/2}}{K_{+}(\mathcal{K})}\left(\frac{K_{+}(k)}{k}\right).
    \label{kutta_2}
\end{equation}
The unsteady Kutta condition demands no singularity in the vicinity of $r\rightarrow 0$, therefore from (\ref{kutta_1}) and (\ref{kutta_2}) we must have
\begin{equation}
     \frac{\mathrm{i}G}{K_{+}(\mathrm{i}\epsilon)} -\frac{\mathrm{i}\mathcal{A}\gamma(\mathcal{K})\mathrm{e}^{-\gamma(\mathcal{K})/2}}{K_{+}(\mathcal{K})}=0.
\end{equation}

%\ccc{Outline of the methods to determine $\epsilon$.} 
The parameter $G$ and the kernel function $K_{+}$ are defined by (\ref{equ:G}) and (\ref{kernel_K}), respectively. To evaluate $\mathcal{A}$, we need to determine the small parameter $\epsilon$, which seems to be an arbitrary value introduced to address the issue of non-convergence. However, $\epsilon$ in fact has a physic-imposed value and we may uniquely determine it by re-evaluating the problem without approximations.

%\ccc{Calculate $\epsilon$ by repeating the Wiener-Hopf technique and deforming the integral path.} 
First, replace the forcing term $GY$ in (\ref{8}) and (\ref{9}) by
\begin{equation}
    F(h)=\frac{1}{2\pi}\int_{-\infty}^{\infty}D_{{\mathrm{o}}}(\lambda)\mathrm{exp}(-\mathrm{i}\lambda h)\mathrm{d}\lambda.
\end{equation}
Then the Wiener-Hopf equation (\ref{Wiener-Hopf_equation}) takes a new form, i.e.
\begin{equation}
     \phi_{+}^{\prime}(k,1/2)+\mathcal{F}(k)=K(k)F_{-}(k).
\end{equation}
Here
\begin{equation}
    \begin{aligned}
          \mathcal{F}(k)%&=\int_{-\infty}^{0}\gamma_{+}F(h)\mathrm{e}^{\mathrm{i}kx}\mathrm{d}x \\ 
    &= \int_{-\infty}^{0} \int_{0}^{\infty} \mathcal{G}(\zeta^{*})\mathrm{exp}(\zeta^{*} x) \mathrm{e}^{\mathrm{i}kx} \mathrm{d}\zeta^{*} \mathrm{d}x,
    \label{equ:2.71}
    \end{aligned}
\end{equation}
where $\zeta^{*}=-\mathrm{i}\lambda$, and the form of $\mathcal{G}(\zeta^{*})$ is shown in Appendix B. If we intend to completely account for the effect of the upstream forcing, we can replace $G$ by $\mathcal{G}(\zeta^{*})$ and $\epsilon$ by $\zeta^{*}$, and then carry out the integration from $0$ to $\infty$. The mathematical proof of this procedure can be found in Appendix C. Thus, it follows that
\begin{equation}
     \frac{G}{K_{+}(\mathrm{i}\epsilon)}=\int_{0}^{\infty} \frac{\mathcal{G}(\zeta^{*})}{K_{+}(\mathrm{i}\zeta^{*})} \mathrm{d}\zeta^{*}.
     \label{80}
\end{equation}
 \label{results}
  \begin{figure}
    \centering
    \includegraphics[width = 0.5\textwidth]{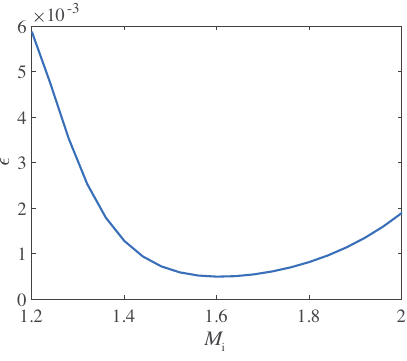}
    \caption{The value of the parameter $\epsilon$ in the range of $1.2<M_{\mathrm{i}}<2$ at the screeching frequency. The frequency is obtained by the original formula (\ref{equ_screech f}) proposed by \citet{19533Powell}.}
 \label{fig:example5}
 \end{figure}
 
%\ccc{Obtain $\epsilon$ using Watson's lemma.} 
The integration on the right-hand side of (\ref{80}) can be evaluated using Watson's Lemma. On the left-hand side of (\ref{80}), given that $\epsilon$ is a small term, $K_{+}(\mathrm{i}\epsilon)$ can be evaluated employing its Taylor expansion. By combining (\ref{K_star_near_zero}) and (\ref{G_zeta_star}), we have
\begin{equation}
   \epsilon=-\frac{\mathrm{i}G^2h^3}{\pi \mathcal{G}^2(0)}.
    \label{equ:epsilon}
\end{equation}
To verify that $\epsilon$ is a small term, we calculate it using (\ref{equ:epsilon}) and plot it in figure \ref{fig:example5}.  The angular frequency of the instability wave is calculated using the original formula (1.1) proposed by \citet{19533Powell}. Figure \ref{fig:example5} shows that within the range of $1.2<M_{\mathrm{i}}<2$, the value of $\epsilon$ initially decreases from its maximal value of $5.9\times 10^{-3}$ at $M_{\mathrm{i}}=1.2$ to $4.9\times 10^{-4}$ at $M_{\mathrm{i}}=1.6$, and then gradually increases to around $1.9\times 10^{-3}$ at $M_{\mathrm{i}}=2$. In the entire Mach number region, $\epsilon$ is small, ensuring the validity of using Taylor expansion in (\ref{K_star_near_zero})  and the omission of $G\epsilon/\omega$  in (\ref{combine2}).

%\ccc{The transfer function.}
 Having obtained all the parameters, the instability wave is found to be
 \begin{equation}
    \phi^{*}(0, 1/2)=\frac{G K_{+}(\mathcal{K})}{\gamma(\mathcal{K})K_{+}(\mathrm{i}\epsilon)}.
 \end{equation}
  This equation enables us to define the important transfer function, i.e.
\begin{equation}
    {H}(\omega, M_{\mathrm{i}}, \nu)=\frac{K_{+}(\mathcal{K})}{\gamma(\mathcal{K})K_{+}(\mathrm{i}\epsilon)},
\end{equation}
 where $K_{+}$ is given by (\ref{kernel_K}), $\mathcal{K}$ by (\ref{4.6}), $\gamma(\mathcal{K})=\sqrt{\mathcal{K}^{2}-\omega^2 M_{\mathrm{o}}^2}$, and $\epsilon$ by (\ref{equ:epsilon}). This transfer function describes the linear mapping between the newly-excited instability waves and the upstream
 propagating acoustic waves. Using this function, both the scattering efficiency and the phase delay between these two waves can be readily examined.
 
 \section{Results}
\label{sec:results}
\begin{figure}
    \centering
    \includegraphics[width = 0.98\textwidth]{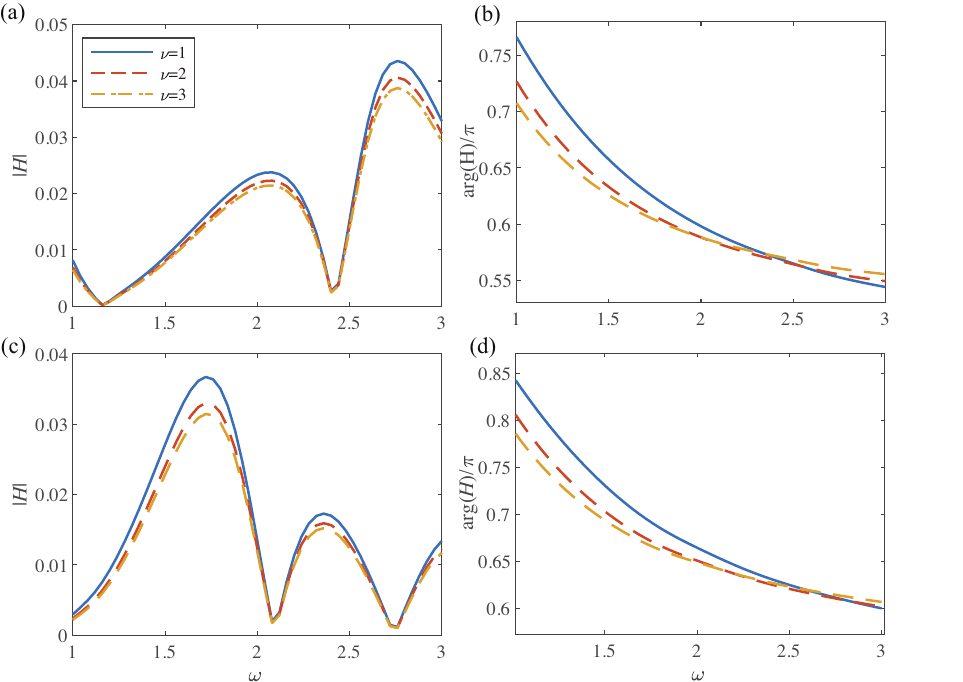}
    \caption{The amplitude and phase of the transfer function $H(\omega,M_{\mathrm{i}},\nu)$. The jet Mach number $M_{\mathrm{i}}$ is 1.3 and 1.5 in (a)-(b) and (c)-(d) respectively.
    The transfer function is examined under three temperature conditions, i.e. $\nu=1$, 2, and 3.}
 \label{fig:example7}
 \end{figure}

%\ccc{Description of the figure regarding the transfer function.} 
We first investigate the behaviour of the transfer function $H$ under two typical screech conditions, i.e. $M_\mathrm{i}=1.3$ and $M_\mathrm{i}=1.5$.
We also consider the effects of the temperature by showing results at three temperature ratios, i.e. $\nu=1$, 2, and 3.
The amplitude, i.e. the scattering efficiency, and the phase of $H$ are shown in figure \ref{fig:example7}. Note that the phase is defined in the range $0\leq\mathrm{arg}(H)< 2\pi$. Considering that the asymptotic expansion yields favourable approximation when $M_\mathrm{i}>1.2$, which corresponds to screech frequency within the range of $1<\omega<3$, only the frequency range from 1 to 3 is examined.
Figures 5(a) and 5(c) show that the scattering efficiency $|H|$ displays multiple peaks and valleys as the frequency varies. In both cases, $|H|$ is less than $5\%$ and decreases to $2\text{\textperthousand}$ at certain frequencies. Note that the temperature has a mild impact on the scattering efficiency. Specifically, as the temperature ratio increases, the scattering efficiency decreases slightly.
Regarding the phase delay, figures 5(b) and 5(d) show a phase delay between around $\pi/2$ to $4\pi/5$, which demonstrates a decreasing trend as the frequency increases. In addition, higher jet temperatures lead to lower phase delays at low frequencies while resulting in slightly higher phase delays at high frequencies.

 \subsection{Phase condition for jet screech}
 %\ccc{Introduce the following sections.} 
Using the transfer function, we can examine the phase condition for the screech feedback loop in this section. In the subsequent analysis, we first focus on the cold jet, which corresponds to the operation conditions commonly employed in experiments and numerical simulations. We then incorporate the thermal influence to account for temperature effects on the screech frequency. 

\begin{figure}
    \centering
    \includegraphics[width = 0.7\textwidth]{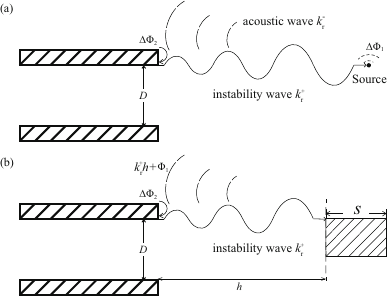}
    \caption{Schematics of the phase condition for the feedback loop of jet screech. (a) Powell's model; (b) this model.}
 \label{fig:example8}
 \end{figure}

%\ccc{The phase condition previously proposed by Powell and Jordan.} 
The original phase condition proposed by~\citet{19533Powell} can be written as 
\begin{equation}
    \frac{N-\mathrm{\Delta} }{f}=\frac{h}{|U^{+}|}+\frac{h}{|U^{-}|},
\end{equation}
where $U^{+}$ and $U^{-}$ represent the nondimensional phase velocity of the downstream-propagating instability waves and the upstream-propagating acoustic waves, respectively. Note that $U^-=1/M_{\mathrm{o}}$ in this study. The integer $N$ denotes the total number of phase cycles contained in the feedback loop, and $\mathrm{\Delta} $ represents a number between 0 and 1 due to the additional phase delay. Note that this phase criterion can be reformulated, as demonstrated by \citet{Jordan_2018} and \citet{NewC_1}, in the following manner, i.e.
\begin{equation}
    (|k_{r}^+|+|k_{r}^-|)h+\mathrm{\Delta} \phi=2\pi N,
    \label{Jordan's Phase condition}
\end{equation}
 where $k_{r}^+$ and $k_{r}^-$ are, respectively, the real part of the wavenumber of the downstream-propagating instability waves and the upstream-propagating acoustic waves. 
 The term $\mathrm{\Delta}\phi$ is defined as $2\pi \mathrm{\Delta}$ and represents the additional phase delay. As illustrated in figure \ref{fig:example8}(a), the instability wave with the wavenumber $k_r^+$ grows spatially from the nozzle lip to the effective source location and interacts with the shock structure. This interaction produces acoustic waves, which propagate to the upstream direction with the wavenumber $k_r^-$ and trigger new instability waves at the nozzle lip. The total phase variations during this feedback loop must be equal to an integer multiplied by $2\pi$. Therefore, the phase condition can be written as (\ref{Jordan's Phase condition}).  As illustrated in figure \ref{fig:example8}(a), the total phase delay $\mathrm{\Delta} \phi$ consists of two components: the phase change occurring during the sound emission ($\mathrm{\Delta} \phi_1$) and that during the jet receptivity ($\mathrm{\Delta} \phi_2$). 
 
 %\ccc{Methods to calculate $k_r^+$.} 
 To predict the screech frequency, the parameters contained in (\ref{Jordan's Phase condition}), such as $k_r^+$, $k_r^-$, and $\Delta \phi$ should be specified, respectively. Regarding $k_{r}^+$, it is commonly assumed that the phase velocity of the jet instability waves $U_c$ is proportional to the velocity of the fully expanded jet flow. Therefore, we can write $k_{r}^+$ as
 \begin{equation}
     \begin{aligned}
        k_{r}^+&=\frac{\omega}{\kappa},
     \end{aligned}
     \label{k_i}
 \end{equation}
 where $\kappa$ typically falls within the range of $0.5$ to $0.8$, and its value depends on factors such as the nozzle shape and the jet Mach number. In circular nozzles, a value of $\kappa=0.7$ may be used~\citep{1973Harper}. However, studies by \citet{X.D.Li} have indicated that $\kappa$ may vary with the jet Mach number. In the case of a rectangular jet, experiments suggested that values between $0.5$ and $0.65$ may be more appropriate compared to $0.7$~\citep{19533Powell,Krothapalli_1986,1994Raman_POF,Panda_1997_rentangular}. Similarly, variations in $\kappa$ due to changes in jet Mach numbers have also been reported~\citep{19533Powell,Panda_1997_rentangular}.
 
 %\ccc{The method to calculate $k_r^+$ in this paper.} 
 Based on experimental measurements and data analysis, \citet{Tam_1996_JoA} proposed an empirical formula to estimate $\kappa$ for rectangular nozzles, i.e.
 \begin{equation}
     \kappa=0.5+0.2\mathrm{e}^{-0.5(\Lambda-1)},
     \label{Tam:empirical}
 \end{equation}
where $\Lambda$ is the aspect ratio of the rectangular nozzle. For a high-aspect-ratio rectangular jet, \citet{Tam_1996_JoA} and \citet{ numerical_directivity_of_rectangular} applied the value of $\kappa=0.55$. In the rest of this paper, if no specific data is available, we use this empirical formula to determine $\kappa$.

 %\ccc{Method to calculate $M_c$.} 
 For $k_{r}^-$, it is straightforward to find that
 \begin{equation}
    |k_{r}^-|=\omega M_{\mathrm{o}}={\omega}M_c/{\kappa}.
    \label{k_r_plus}
 \end{equation} 
 Combining (\ref{equ:n}), (\ref{Jordan's Phase condition}), (\ref{k_i}), and (\ref{k_r_plus}) yields
 \begin{equation}
     f=\frac{2\pi N-\mathrm{\Delta} \phi}{2 \pi n}\frac{\kappa}{s(1+M_c)}.
     \label{equ:87}
 \end{equation}
 Note that, the additional phase delay $\mathrm{\Delta}\phi$ is typically assumed to be $0$ in previous studies~\citep{Jordan_2018,xiangru_Li_2023}. The integer $N$, according to the findings of \citet{2017_jfm_sources}, is often equal to the shock cell number $n$ for circular nozzles. We assume that this relation holds for the two-dimensional jet under consideration. By setting $\mathrm{\Delta} \phi=0$ and $N=n$, equation (\ref{equ:87}) can be simplified to (\ref{equ_screech f}), which represents the original frequency prediction model proposed by~\citet{19533Powell}. Note that (\ref{equ_screech f}) is derived under the assumption that the sound intensity in the upstream direction reaches its maximum, thus corresponding to the constructive interference condition. We can see that the frequency prediction formula obtained from the phase condition coincides with (\ref{equ_screech f}) when $\mathrm{\Delta} \phi=0$ and $N=n$.

% \ccc{A new phase condition.} 
As we can see from figure \ref{fig:example8}(a), an effective screech source is assumed in early research on the phase condition. However, whether the screech source is distributed or localized is still open to debate. As illustrated in figure \ref{fig:example8}(b), in our previous study~\citep{MyOwn_2}, the acoustic wave is generated by the interaction between the instability wave and the distributed shock cells. Consequently, the sound source is distributed across various shock structures, e.g. the fourth and fifth shock structures. Therefore, we cannot pinpoint a specific point source. Nevertheless, without an effective source location, we can still establish a phase condition in this model, i.e.
 \begin{equation}
     k_r^{+} h+\Phi_1+{\Delta} \Phi_2=2\pi N^*.
     \label{equ:3.8}
 \end{equation}
 Here, $h$ denotes the distance between the nozzle lip and the left boundary of the source region as shown in figure \ref{fig:example8}(b). $\Phi_1$ represents the phase of the resulting acoustic wave at the nozzle lip, which is generated from the interaction between the shock structures and an instability wave whose phase is set to be 0 at the left boundary of the source region. The actual phase of the acoustic wave at the nozzle lip, after taking account of the phase variation due to instability propagation, would be $k^{+}_r h+\Phi_1$, as shown in figure \ref{fig:example8}(b). $\Delta\Phi_2$ denotes the phase change during the jet receptivity, which is the same as that defined in figure \ref{fig:example8}(a). Similar to (\ref{Jordan's Phase condition}), the total phase variations during the feedback loop should be equal to an integer multiplied by $2\pi$. Therefore, we can write the phase condition as (\ref{equ:3.8}), where a new integer $N^*$ is introduced. We expect $N^{*}< N$ because the actual phase variation between the left boundary of the source region and the nozzle lip is probably more than $2\pi$. However, only $\Phi_1$ is included, which is less than $2\pi$ under several operation conditions. We find that $N^{*}=N-1$ is often appropriate.  From (\ref{k_i}) and (\ref{equ:3.8}), we can obtain
 \begin{equation}
     f=\frac{2\pi N^*-\Phi_1-\mathrm{\Delta} \Phi_2}{2\pi ns}\kappa.
     \label{equ:new_phase_condition}
 \end{equation}
  Equipped with ${\Delta}\Phi_2$ obtained from the transfer function and $\Phi_1$ obtained from \citet{MyOwn_2}, we can compare this model prediction with Powell's model prediction and the experimental measurements.

\begin{table}
    \begin{center}
    \begin{tabular}{lcccccc}
        Cases   & $\kappa$   & $\Lambda$ & $M_d$ &  $\nu$   & $N$ &$N^*$\\[3pt]
         \citet{19533Powell}   &{0.6} &  {5.83} &  {1} &  {1} &  {4} & {3}\\
         \citet{1994Raman_POF}  &  {0.54} & {9.63} & {1} & {1.42}& {4} &  {3}\\
         \citet{Panda_1997_rentangular}  &  {0.65} &  {5} & {1} & {1} & {4} &  {3}\\
         \citet{2003Alkislar}  &  {0.63} &  {4}& {1.44} & {1} & {4} &  {3}\\
    \end{tabular}
    \caption{The operation conditions used in several experiments.}
    \label{tab:2}
    \end{center}
\end{table}

\begin{figure}
    \centering
    \includegraphics[width = 0.92\textwidth]{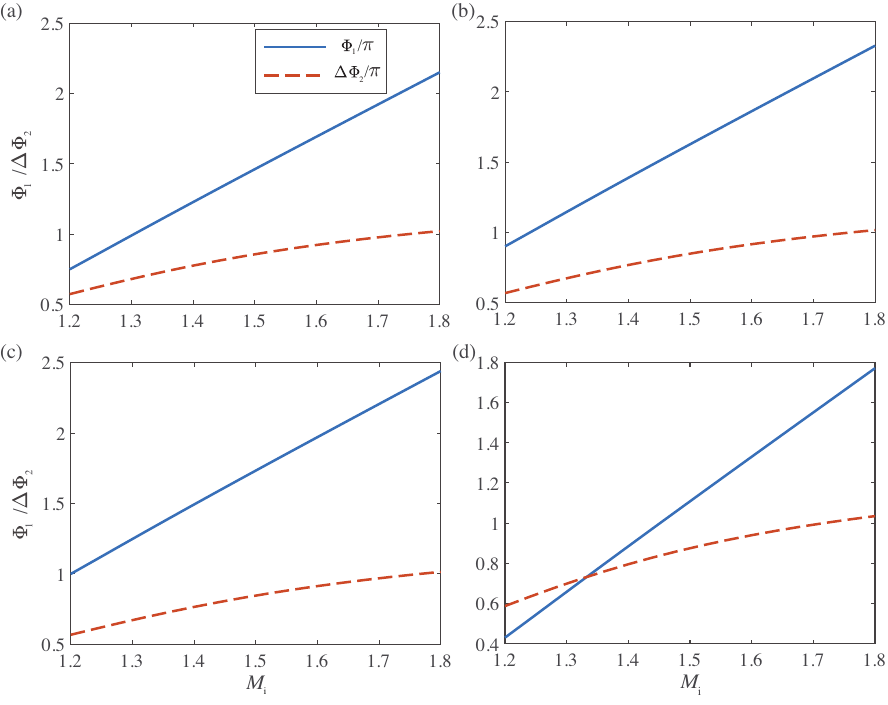}
    \caption{The phase changes $\Phi_1$ and $\Delta \Phi_2$ at the screeching frequency. The operation conditions used in figures with labels (a)-(d) are the same as those used in \citet{1952Powell}, \citet{1994Raman_POF}, \citet{Panda_1997_rentangular}, and \citet{2003Alkislar}, respectively.}
 \label{fig:examplep_angle}
 \end{figure}

 The operation conditions used in four experiments in the open literature, including the aspect ratio of the rectangular nozzle ($\Lambda$), the designed Mach number of the nozzle ($M_d$), and the temperature ratio ($\nu$) are listed in table \ref{tab:2}. In these experiments, the convection velocities of the instability wave were all measured. The constants $\kappa$ obtained from these experiments are listed in table \ref{tab:2}. Note that in \citet{Panda_1997_rentangular}, the constant $\kappa$ was found to increase as $M_\mathrm{i}$ increased, from 0.6 at low $M_\mathrm{i}$ to 0.7 at high $M_\mathrm{i}$. We use an average value of $0.65$ to initiate the computation. In addition, the screech source region was reported to be around the fourth shock structure from experimental observations, thus we set $N=4$ and $N^*=3$ in the four cases. To calculate the shock spacing $s$ in (\ref{equ:87}) and (\ref{equ:new_phase_condition}), we use the model proposed by~\citet{Tam_1988jsv}, where $s$ can be expressed by
\begin{equation}
    s=2\sqrt{M_\mathrm{i}^2-1}\left[\left(\frac{{1}}{{{W}}}\right)^2+1\right]^{-\frac{1}{2}}.
    \label{equ:tam_shockspacing}
\end{equation}
Here  ${{W}}$  represent the nondimensional height and width of the fully expanded jet flow. $1/{W}$ is related to the jet aspect ratio ($\Lambda$), the Mach number of the fully expanded jet flow ($M_\mathrm{i}$), and the designed Mach number ($M_d$). The method to calculate $1/{W}$ can be found in \citet{Tam_shock_space}. 

 \begin{figure}
    \centering
    \includegraphics[width = 0.92\textwidth]{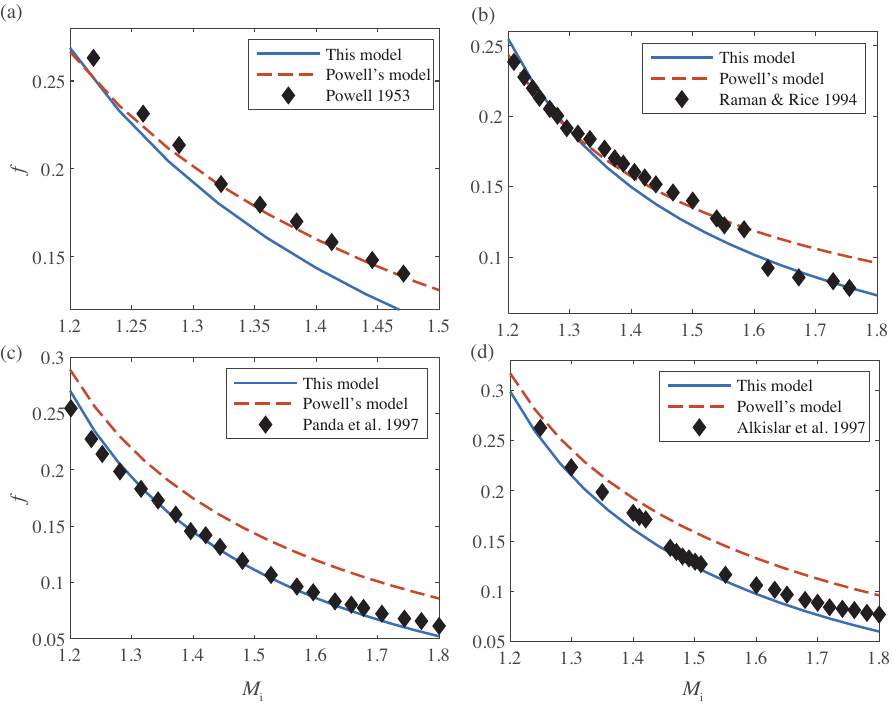}
    \caption{Comparison between Tam's model prediction, this model prediction, and the experimental data. The solid blue curve is calculated by Powell's model. The dashed red line is obtained using this model. Experimental data in figures with label (a)-(d) are measured by \citet{1952Powell}, \citet{1994Raman_POF}, \citet{Panda_1997_rentangular}, and \citet{2003Alkislar}, respectively.}
 \label{fig:examplep_fre}
 \end{figure}
The phase $\Phi_1$ and the phase delay $\Delta \Phi_2$ can be readily calculated if $f$ is provided. However, $f$ depends on  $\Phi_1$ and $\Delta \Phi_2$ via (\ref{equ:new_phase_condition}). So
to determine $\Phi_1$ and $\Delta \Phi_2$ in (\ref{equ:new_phase_condition}), we use an iterative scheme as follows. We first employ the operation conditions presented in table~\ref{tab:2} and calculate the screech frequency using Powell's original formula (\ref{equ_screech f}). Subsequently, the obtained frequency is substituted into this model to determine $\Phi_1$ and $\Delta \Phi_2$. A new screech frequency can then be predicted using (\ref{equ:new_phase_condition}). This new frequency can be used to obtain an updated $\Phi_1$ and $\Delta \Phi_2$. The procedures are repeated until a converged frequency is obtained. We first show the phase changes in figure \ref{fig:examplep_angle}. As can be seen, both $\Phi_1$ and $\Delta \Phi_2$ increase as the jet Mach number increases. $\Phi_1$ increases in a nearly linear manner with respect to $M_\mathrm{i}$, while $\Delta \Phi_2$ varies slowly as $M_\mathrm{i}$ changes, increasing from nearly $\pi/2$ at low Mach numbers to $\pi$ as $M_\mathrm{i}$ increases to 1.8.

Equipped with the $\Phi_1$ and $\Delta \Phi_2$, we can calculate $f$ from (\ref{equ:new_phase_condition}). Comparison between this model prediction, Powell's model prediction, and the experimental data are shown in figure \ref{fig:examplep_fre}. It can be seen that this model predictions agree well with the experimental data in figures \ref{fig:examplep_fre}(c)-(d), while Powell's model predictions seem to show an overprediction compared to the experiments. Note that in figure \ref{fig:examplep_fre}(c), setting $\kappa=0.5$ can lead to favourable agreement between Powell's model prediction and the experimental data, as shown by \citet{numerical_directivity_of_rectangular}, whereas using the convection velocity directly obtained from the experimental data leads to an overprediction. In figure \ref{fig:examplep_fre}(a), on the other hand, this model underpredicts the screech frequency, while Powell's model prediction agrees well with the experimental data. Note, however, if the constant $\kappa$ is assumed to be $0.7$, Powell's model overpredicts the screech frequency compared with the experimental data, as shown by \citet{Tam_1988jsv}, while this model prediction would show good agreement with the experimental data. Whether this underprediction is due to an inaccurate $\kappa$ is not yet clear and therefore requires further scrutiny. In figure \ref{fig:examplep_fre}(b), both of the two models provide favourable results. Specifically, Powell's model prediction agrees well with the experimental data when $M_\mathrm{i}<1.6$, while this model prediction agrees well with the experimental data at low Mach numbers and high Mach numbers.
%if the aspect ratio is assumed to be large enough in figure \ref{fig:examplep_fre}(a), Powell's model prediction agrees well with the experimental data, as shown by \citet{Tam_1988jsv}, but when considering the effect of the small aspect ratio, overprediction occurs between the model prediction and the experiment results.
 Generally speaking, the model prediction demonstrates favourable agreement with the experimental data~\citep{1994Raman_POF,Panda_1997_rentangular,2003Alkislar}, suggesting that considering the additional phase delay may rectify the overprediction of the classical model.

%\ccc{Temperature effects on the phase condition. Introduce one approach in detail.} 
In the final part of this section, we consider the thermal influence on the phase condition, which can be accounted for in two ways. Tranditional model of the thermal effect is through the modification of $U_c$, as shown by (\ref{M_c}). This effect was previously considered by~\citet{1986Tam_Proposed} for circular jet flow. They found that under 
different temperature ratios, the screech frequency can be predicted by
 \begin{equation}
     f=0.67(M_{\mathrm{i}}^2-1)^{-1/2}\left[1+0.7M_{\mathrm{i}}\left(1+\frac{\gamma-1}{2}M_{\mathrm{i}}^2\right)^{-1/2}\nu^{1/2}\right]^{-1}.
     \label{equ:89}
 \end{equation}
In the rectangular jet flows, a similar formula may be obtained from (\ref{equ:87}) and (\ref{equ:tam_shockspacing}), assuming $N=n$, $\mathrm{\Delta} \phi=0$, and $\kappa=0.7$, i.e.
 \begin{equation}
     f=0.35(M_{\mathrm{i}}^2-1)^{-1/2}\left[1+0.7M_{\mathrm{i}}\left(1+\frac{\gamma-1}{2}M_{\mathrm{i}}^2\right)^{-1/2}\nu^{1/2}\right]^{-1}\left[\left(\frac{1}{{W}}\right)^2+1\right]^{\frac{1}{2}}.
     \label{equ:90}
 \end{equation}

\begin{table}
    \begin{center}
    \begin{tabular}{lcccccc}
        Cases   & $\kappa$   & $\Lambda$ & $M_d$ &  $\nu$   & $N$ &$N^*$\\[3pt]
         1  &{0.8} &  {2} &  {1.5} &  {1} &  {4} & {3}\\
         2  &  {0.73} & {2} & {1.5} & {2}& {4} &  {3}\\
        3 &  {0.64} &  {2} & {1.5} & {3} & {4} &  {3}\\
        
    \end{tabular}
    \caption{The operation conditions used in \citet{Gojon_2019}.}
    \label{tab:3}
    \end{center}
\end{table}

 \begin{figure}
    \centering
    \includegraphics[width = 0.5\textwidth]{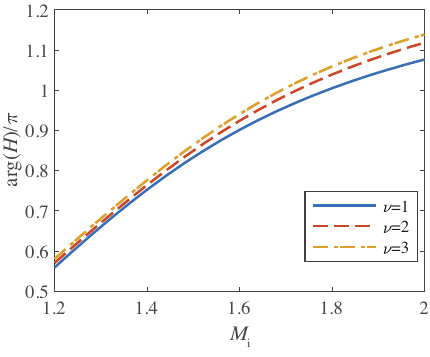}
    \caption{The phase delay between the upstream forcing and newly-excited instability wave at the screeching frequency. The parameter $\kappa$ takes the value of 0.7 in the three cases and the angular frequency is calculated using (\ref{equ_screech f}).}
 \label{fig:examplehotphase}
 \end{figure}
 
 %\ccc{Interpreting the above two equations.} 
 Note that the difference between (\ref{equ:89}) and (\ref{equ:90}) arises from the distinct values of the shock spacing in circular and rectangular jet flows, which can be predicted by Pack's model~\citep{1950Pack,Tam_1988jsv}, i.e. $s={\pi} \sqrt{M_{\mathrm{i}}^2-1}/2.4048$ for circular nozzles while $s=2\sqrt{M_{\mathrm{i}}^2-1}/\sqrt{{1}/{{W}^2}+1}$ for rectangular cases.
  \begin{figure}
    \centering
    \includegraphics[width = 0.5\textwidth]{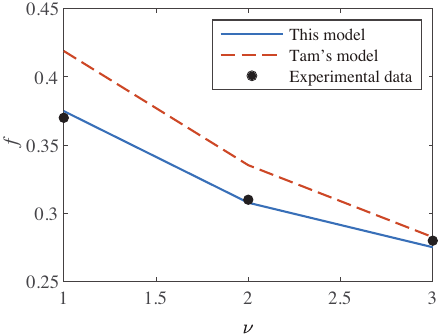}
    \caption{Comparison of Tam's model prediction, this model prediction, and the experimental data~\citep{Gojon_2019}. The dashed red curve is calculated by Tam's model. The solid blue line is obtained using this model. Filled circle points represent experimental data from~\citet{Gojon_2019}.}
 \label{fig:example11}
 \end{figure}
 
 %\ccc{Introduce the other approach.} 
 In this model, from figure \ref{fig:example7}, we can see that the phase delay is also changed by the temperature ratios. Therefore, to predict the screech frequency more accurately, the changes in the phase delay and the convection velocity $U_c$ should be both accounted for. %Different temperature ratios result in varying values of $\sigma$, which can be calculated using (\ref{M_c}), i.e. 
% \begin{equation}
  %   \sigma=\dfrac{1}{1+\dfrac{\gamma-1}{2}M_{\mathrm{i}}^2}\nu.
 %\end{equation}
%The variations in $\sigma$ leads to changes in the phase delay of the transfer function. 
To quantitatively evaluate the thermal effects, we take three temperature ratios, i.e. $\nu=1$, $2$, and $3$ into consideration. We calculate the transfer function at the screeching frequency under the three temperature ratios, the phase delays of which are shown in figure \ref{fig:examplehotphase}. Note that the screech frequency is calculated via (\ref{equ_screech f}) with the constant $\kappa=0.7$.
We can see that under the three temperature ratios phase delays increase as the jet Mach number increases. In addition, higher jet temperatures lead to larger phase delays ($\Delta \Phi_2$) across the entire range of Mach numbers, which will, according to (\ref{equ:new_phase_condition}), result in a lower frequency prediction. Note that this appears to contradict figure \ref{fig:example7}, but this apparent contradiction results from the fact that we are considering the phase delay at the screech frequency for each $M_\mathrm{i}$ and $\nu$, whereas figure \ref{fig:example7} considers the phase delay for a given $M_\mathrm{i}$ at the same frequency.

 %\ccc{Compare this model prediction, Tam's model prediction and numerical data.} 
 To quantitatively validate (\ref{equ:new_phase_condition}), we use the data from~\citet{Gojon_2019}, where overexpanded jet flows under different temperature conditions were simulated using LES. The convection velocity of the instability waves was numerically obtained. They compared the frequency prediction from Tam's model~\citep{Tam_1988jsv} to the reported experimental results~\citep{experimentaldata}. ``An overestimation of about $10\%$" \citep{Gojon_2019} compared to the experimental data was reported. This overprediction may be rectified if the calculated $\Delta \Phi_2$ is considered. To obtain the screech frequency prediction, we employ the data of the convection velocity from \citet{Gojon_2019} and evaluate (\ref{equ:new_phase_condition}) under the three different temperature ratios. The operation conditions are the same as those used in \citet{Gojon_2019} and are listed in table \ref{tab:3}. The comparison is shown in figure \ref{fig:example11}. These results indicate that if the additional phase delay is included, the predicted screech frequency agrees much better with the numerical simulation. Specifically, the overprediction from the classical model is rectified when the phase delay is considered.

 \section{Conclusion}
\label{sec:summary}
This paper investigates the jet receptivity occurring at the nozzle lip within the screech feedback loop. The scattered field due to the upstream-propagating acoustic waves is derived using the Wiener-Hopf method, with the kernel decomposed through matched asymptotic expansions. To determine the newly-triggered instability wave, we impose the unsteady kutta condition by eliminating the emerging singularities in the scattered field. This results in a dispersion relation that governs the newly-excited instability waves. The transfer function between the newly-excited instability waves and the upstream forcing is subsequently obtained. The amplitude and phase of the transfer function are then discussed in detail.

The result shows that the scattering efficiency $|H|$ displays multiple peaks and valleys as the frequency increases. In addition, it is less than $5\%$ under each operation condition that we consider.  Moreover, the phase delay between the instability wave and the upstream forcing generally decreases as the frequency increases. Regarding the thermal influence on the transfer function, it is found that the scattering efficiency decreases slightly as the temperature ratio increases from 1 to 3, while higher jet temperatures can lead to lower phase delays at low frequencies but result in slightly higher phase delays at high frequencies. 

Including this additional phase delay, we invoke the phase condition to predict the screech frequency. The new prediction demonstrates favourable agreement with the experimental data for cold supersonic jets. In particular, the overprediction of screech frequencies from classical models appears to be rectified using the present model, except for the case of Powell's measurement. In addition, we examine the effects of the temperature ratio on the screech frequency by including the additional phase delay, the result of which shows that the screech frequency prediction improves significantly.

Our future work would focus on the scattering efficiency obtained by the transfer function, which is not thoroughly investigated in this paper due to a lack of direct validation from experimental or numerical data in the open literature. Moreover, the frequency range under consideration is relatively small due to the requirement of the matched asymptotic expansions. We consider to use new strategies to overcome the limitations in our future study.

 %one near $\pi/2$, the other close to $3\pi/2$. The phase difference of the two trajectories equals $\pi$. 
 %For each case, although the amplitude of instability is continuous as $M_{\mathrm{i}}$ changes, the argument part shows an obvious staging characteristic. For instance, as $n=3$, the staging occurs at around $M_{\mathrm{i}}=1.4$ and $M_{\mathrm{i}}=1.7$, with an abrupt increase or decrease of $\pi$.

\appendix

\section{}
\label{appendix b}
 \begin{figure}
    \centering
    \includegraphics[width = 0.95\textwidth]{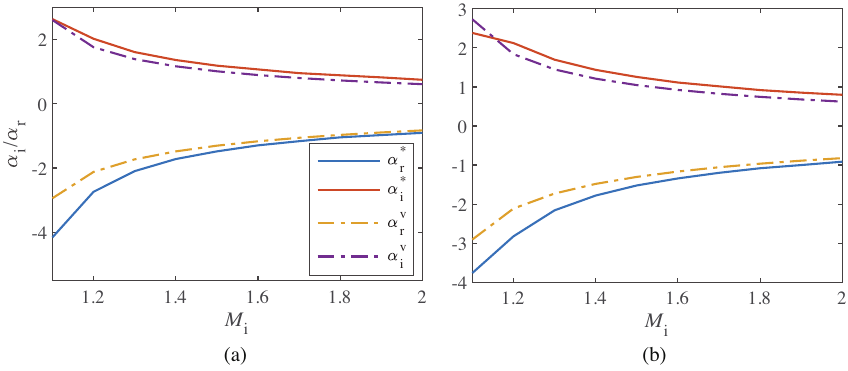}
    \caption{Comparison between the wavenumber numerically obtained by the dispersion relation and that calculated through the asymptotic expansions. The temperature ratio $\nu$ equals 3 and 5 in (a) and (b), respectively.}
 \label{fig:example9}
 \end{figure}

 Following the procedure of Ferrari's formula, zeros of the polynomial can be determined. Define a parameter $y_a$ as
 \begin{equation}
     y_a=\sqrt[3]{c_2+\sqrt{c_3}}+\sqrt[3]{c_2-\sqrt{c_3}},
 \end{equation}
 where
 \begin{gather}
    c_2=a_1^2/2,\\
    c_3=c_2^2-64a_0^3/27.
 \end{gather}
 The zeros $q_1,q_5,q_2,q_6$ take the following forms, i.e.
 \begin{equation}
    q_{1,5}=-\frac{1}{2}\sqrt{y_a}\pm\frac{1}{2}\sqrt{-y_a+\frac{2a_1}{\sqrt{y_a}}
    },
 \end{equation}
 \begin{equation}
    q_{2,6}=\frac{1}{2}\sqrt{y_a}\pm\frac{1}{2}\sqrt{-y_a-\frac{2a_1}{\sqrt{y_a}}
    }.
 \end{equation}
 Zeros $q_3, q_7, q_4, q_8$ take similar forms, i.e.
 \begin{equation}
    q_{3,7}=-\frac{1}{2}\sqrt{y_a^{*}}\pm\frac{1}{2}\sqrt{-y_a^{*}-\frac{2a_1}{\sqrt{y_a^{*}}}
    },
    \label{q3_star}
 \end{equation}
 \begin{equation}
    q_{4,8}=\frac{1}{2}\sqrt{y_a^{*}}\pm\frac{1}{2}\sqrt{-y_a^{*}+\frac{2a_1}{\sqrt{y_a^{*}}}
    },
    \label{4.6}
 \end{equation}
 where
 \begin{equation}
     y_a^{*}=\sqrt[3]{c_2+\sqrt{c_2^2+c_3^{*}}}+\sqrt[3]{c_2-\sqrt{c_2^2+c_3^{*}}},
 \end{equation}
 \begin{equation}
     c_3^{*}=c_2^2-64a_0^3/27.
 \end{equation}
It can be shown, by some algebra, that when the angular frequency $\omega>0.05$ and  $1<M_{\mathrm{i}}<2$, $c_3>0$ and $c_3^{*}>0$. In addition, $y_a^{*}$ can be shown to be a real number. 

With a negative real part and a positive imaginary part, $q_3 S^{4/3}$ ($S$ is defined in section \ref{sec:B}) corresponds to the wavenumber of the jet instability of the sinuous mode, which is defined as $\mathcal{K}$ in (\ref{equ:instability}). The original polynomial $R(q)$ has four zeros denoted as $q_{2,6}$ (in the right half-plane $\mathrm{Re}(q)>0$) and $q_{3,7}$ (in the left half-plane $\mathrm{Re}(q)<0$). The additional zeros are due to the increase in the polynomial order.

To further validate the decomposition, we present the additional comparison between the wavenumber numerically obtained by the dispersion relation and that calculated through the asymptotic expansions using different temperature ratios. We see that for the heated jets under consideration ($\nu=3$ or $\nu=5$) the agreement between the two wavenumbers remains satisfactory as $M_{\mathrm{i}}>1.2$. 
\section{}
\label{appendix D}
We first provide the detailed derivation from (\ref{ln form}) to (\ref{R+R-}). Substracting the left-hand side and right-hand side of (\ref{ln form}) with $\sum_{n=1}^8 1/2(q-q_n)$, the left-hand side part can be directly written as
\begin{equation*}
    \frac{\mathrm{d}}{\mathrm{d} q} \ln \left(\frac{R_{+}(q)}{\prod_{n=5}^8\left(q-q_n\right)^{1/2}}\right)+\frac{\mathrm{d}}{\mathrm{d} q} \ln \left(\frac{R_{-}(q)}{\prod_{n=1}^4\left(q-q_n\right)^{1/2}}\right).
\end{equation*}
 Using partial fraction decomposition, we can assume that the right-hand side part takes the following form, i.e.
\begin{equation}
    \frac{(4q^3+a_1 |q|/q)(q^4-a_1|q|+a_0)}{(q^4+a_0)^2-a_1^2q^2}-\sum_{n=1}^8\frac{1}{2(q-q_n)}=\sum_{n=1}^8 \xi_n \frac{1}{q-q_n}+\xi_9 \frac{1}{q-q_9},
    \label{c1}
\end{equation}
where $q_9=0$. The undetermined coefficient $\xi_n$ can be calculated by multiplying the left-hand and right-hand side of (\ref{c1}) with the corresponding $q-q_n$ and letting $q=q_n$ subsequently.

In what follows, we provide the detailed derivation from (\ref{R+R-}) to (\ref{Rplus}). Using the addictive decomposition (\ref{equ:addictive decomposition}) and the identity (\ref{equ:relation1}), the right-hand side of (\ref{R+R-}) can be rewritten as
\begin{equation}
    \sum_{n=1}^9\xi_n\frac{q_n+\dfrac{\mathrm{i}q_n}{\pi}\ln_+ q-\dfrac{\mathrm{i}q_n}{\pi}\ln_- q}{q-q_n},
    \label{c2}
\end{equation}
which can be reorganized as
\begin{equation}
    \sum_{n=1}^9\xi_n\frac{\left(q_n+\dfrac{\mathrm{i}q_n}{\pi}\ln_+ q_n-\dfrac{\mathrm{i}q_n}{\pi}\ln_- -q_n\right)+\dfrac{\mathrm{i}q_n}{\pi}\ln_+ q/q_n-\dfrac{\mathrm{i}q_n}{\pi}\ln_- -q/q_n}{q-q_n}.
    \label{c3}
\end{equation}
Using the identity (\ref{equ:relation2}), we can reduce (\ref{c3}) to
\begin{equation*}
    \sum_{n=1}^9\xi_n\frac{\dfrac{\mathrm{i}q_n}{\pi}\ln_+ q/q_n}{q-q_n}-\sum_{n=1}^9\xi_n\frac{\dfrac{\mathrm{i}q_n}{\pi}\ln_- q/q_n}{q-q_n}-\sum_{n=1}^9\xi_n \frac{q_n}{q-q_n}.
\end{equation*}
Considering that $R_+ (q)$ and $R_- (q)$ should be analytic and non-zero in the upper and lower half-planes respectively, we can properly write $R_+ (q)$ as the form shown in (\ref{Rplus}).

\section{}
 \label{appendix c}
To calculate the acoustic wave propagating just outside the jet flow, i.e. $\phi_{\mathrm{in}}(x,1/2)$, it is convenient to deform the integration path in (\ref{initial_integraion}) along the branch cut of $\gamma_1$ at $k=\omega M_{\mathrm{o}}$. In this way, the difference in the value of $D_{\mathrm{o}}(\lambda)$ on two different sides of the branch cut only comes from $\gamma_{+}$. $\phi_{\mathrm{in}}(x,1/2)$ can be thus expressed as
\begin{equation}
        \phi_{\mathrm{in}}(x,1/2)=\frac{\mathrm{i}}{2\pi}\mathrm{e}^{-\mathrm{i}M_{\mathrm{o}}\omega x }\int_{0}^{+\infty}[D_{{\mathrm{o}}}^{*}(M_{\mathrm{o}}\omega+\mathrm{e}^{\mathrm{i}\pi/2}\zeta)-D_{{\mathrm{o}}}^{*}(M_{\mathrm{o}}\omega+\mathrm{e}^{\mathrm{i}3\pi/2}\zeta)]\mathrm{e}^{\zeta x}\mathrm{d}\zeta,
\end{equation}
where $D^{*}(\zeta)=D_{\mathrm{o}}(\zeta)/\mathrm{exp}(\gamma_{+}/2)$, and $\lambda=\omega M_{\mathrm{o}}+\mathrm{i}\zeta$.
It is straightforward to see that only $\eta(\lambda)$ contains $\gamma_{+}$ (the definition of $\eta(\lambda)$ can be found in \citet{MyOwn_2}). We can define a new function $E_{{\mathrm{o}}}(\zeta)$ as follows:
\begin{equation}
    E_{{\mathrm{o}}}(\zeta)=D_{{\mathrm{o}}}^{*}(\zeta)\eta(\zeta).
\end{equation}
$\phi_{\mathrm{in}}(x,1/2)$ can be then reorganized as
\begin{equation}
    \phi_{\mathrm{in}}(x,1/2)=\frac{-\mathrm{i}}{\pi}\mathrm{e}^{-\mathrm{i}M_{\mathrm{o}}\omega x }\int_{0}^{+\infty}E_{{\mathrm{o}}}(\zeta)\frac{\omega^3\coth(\gamma_{-}/2)\gamma_{-}}{\omega^{4}\coth^{2}(\gamma_{-}/2)\gamma_{-}^2-(\omega+\lambda)^{4}\mathrm{\Delta}^{2}(\zeta)\frac{M_{\mathrm{i}}^4}{M_{\mathrm{o}}^4}} \mathrm{e}^{\zeta x} \mathrm{d}\zeta,
\end{equation}
where
\begin{equation}
    \mathrm{\Delta}(\zeta)=\sqrt{\mathrm{e}^{\pi\mathrm{i}/2 }\zeta(\mathrm{i}\zeta+2M_{\mathrm{o}}\omega)}.
\end{equation}
Letting $\zeta^{*}=\zeta-\mathrm{i}M_{\mathrm{o}}\omega=-\mathrm{i}\lambda$ and performing a coordinate transformation from $x=0$ to $x=-h$, we can write the function $\mathcal{G}(\lambda)$ in (\ref{equ:2.71}) as 
\begin{equation}
    \mathcal{G}(\zeta^{*})=\frac{-\mathrm{i}}{\pi}E_{\mathrm{o}}(\zeta^{*})\frac{\omega^3\coth(\gamma_{-}/2)\gamma_{-}}{\omega^{4}\coth^{2}(\gamma_{-}/2)\gamma_{-}^2-(\omega+\lambda)^{4}\mathrm{\Delta}^{2}(\zeta^{*})\frac{M_{\mathrm{i}}^4}{M_{\mathrm{o}}^4}} \mathrm{e}^{-\zeta^{*}h}.
    \label{b5}
\end{equation}

It can be proved that as $\zeta^{*}\rightarrow 0$, $\mathcal{G}(\zeta^{*})$ takes the following asymptotic forms, i.e.
\begin{equation}
\mathcal{G}(\zeta^{*})=\mathcal{G}(0)+\mathcal{C}\zeta^{*}+O(\zeta^{*^2}).
\label{G_zeta_star}
\end{equation}
Here, $\mathcal{G}(0)$ is what we aim to find to determine $\epsilon$ in (\ref{equ:epsilon}), which can be calculateed by letting $\zeta^*=0$ in (\ref{b5}), while $\mathcal{C}$, a coefficient of order $O(1)$, is not presented here for clarity.

In what follows, we provide a proof of the relation (\ref{80}), which is essential to uniquely  determine $\epsilon$. The approximated form of the upstream forcing can be written as
\begin{equation}
    G\mathrm{e}^{\epsilon x},
    \label{equC:appro}
\end{equation}
while without an approximation, the forcing reads
\begin{equation}
    \int_{0}^{\infty} \mathcal{G}(\zeta^{*})\mathrm{exp}(\zeta^{*} x) \mathrm{d}\zeta^{*}.
    \label{equC:full_expression}
\end{equation} 
These two expressions should be virtually the same when $x<0$.
Multiplying (\ref{equC:appro}) and (\ref{equC:full_expression}) with $\mathrm{e}^{\mathrm{i}kx}/K_+(k)$ and integrating each expression with respect to $x$ from $-\infty$ to $\infty$, we can find that (\ref{equC:appro}) and (\ref{equC:full_expression}) can be respectively evaluated as
\begin{equation}
    \frac{2\pi G\delta(\mathrm{i}k+\epsilon)}{K_+(k)} \quad \mathrm{and} \quad  \frac{2\pi\int_{0}^{\infty} \mathcal{G}(\zeta^{*})\delta(\mathrm{i}k+\zeta^*) \mathrm{d}\zeta^{*}}{K_+(k)}.
\end{equation}
Subsequently, integrating each expression with respect to $k$ from $-\infty$ to $\infty$ yields (\ref{80}).
\section*{Acknowledgments}
The first author B. Li gratefully acknowledge Y. Ye for his helpful advice on numerical integration.
The authors wish to gratefully acknowledge the financial support from Laoshan Laboratory under
the grant number of LSKJ202202000.
\bibliographystyle{jfm}
% Note the spaces between the initials
\bibliography{jfm-instructions}
\end{document}